\documentclass[journal]{IEEEtran}

\def\LB{\left(}         
\def\RB{\right)}        

\newfont{\bbb}{msbm10 scaled 500}

\newfont{\bb}{msbm10 scaled 1100}

\newcommand{\RR}{\mbox{\bb R}}

\newcommand{\bv}{{\bf b}}

\newcommand{\fv}{{\bf f}}

\newcommand{\kv}{{\bf k}}

\newcommand{\pv}{{\bf p}}

\newcommand{\yv}{{\bf y}}
\newcommand{\zv}{{\bf z}}

\newcommand{\Bm}{{\bf B}}
\newcommand{\Cm}{{\bf C}}
\newcommand{\Dm}{{\bf D}}

\newcommand{\Gm}{{\bf G}}

\newcommand{\Id}{{\bf I}}

\newcommand{\Km}{{\bf K}}

\newcommand{\Mm}{{\bf M}}

\newcommand{\deltav}{\hbox{\boldmath$\delta$}}

\newcommand{\epsilonv}{\hbox{\boldmath$\epsilon$}}

\newcommand{\thetav}{\hbox{\boldmath$\theta$}}

\newcommand{\omegav}{\hbox{\boldmath$\omega$}}

\renewcommand{\arg}{{\hbox{arg}}}

\usepackage{cite}
\usepackage[english]{babel}
\usepackage{algorithm}
\usepackage[utf8]{inputenc}
\usepackage[noend]{algpseudocode}
\usepackage{amsmath,amssymb,amsfonts}
\usepackage{verbatim}
\usepackage{amsmath}
\usepackage{amsmath,amssymb,amsthm}
\usepackage{array,booktabs}
\usepackage{multirow}
\usepackage{graphicx}
\usepackage{textcomp}
\usepackage{caption}
\usepackage{soul,xcolor}
\usepackage{comment}
\usepackage{multirow}
\usepackage{color, colortbl}
\usepackage{subfigure}
\usepackage{hyperref}
\usepackage{float}
\usepackage{nomencl}
\usepackage{wasysym}
\usepackage{soul}

\usepackage{cite}
\usepackage{amsmath,amssymb,amsfonts}
\usepackage{textcomp}
\usepackage{comment}
\usepackage{color, soul}
\usepackage{enumitem}
\setlist[itemize]{leftmargin=0pt}
\usepackage{setspace}
\usepackage{multirow}
\usepackage{nomencl}
\addto\captionsenglish{}

\newcommand{\derip}[1]{\ensuremath{{\frac {\partial #1} {\partial M_g} } } }

\DeclareMathOperator*{\argmax}{arg\,max}

\usepackage{tikz}

\newtheorem{lemma}{\bf{Lemma}}

\newtheorem{theorem}{\bf{Theorem}}

\begin{document}

\title{Load-Altering Attacks Against Power Grids under COVID-19  Low-Inertia Conditions}

\author{Subhash Lakshminarayana,~\IEEEmembership{Senior Member,~IEEE,}
        Juan Ospina,~\IEEEmembership{Member,~IEEE,} and Charalambos~Konstantinou,~\IEEEmembership{Senior~Member,~IEEE}
        \thanks{{Subhash Lakshminarayana is with the School of Engineering, University of Warwick, UK ( E-mail:  subhash.lakshminarayana@warwick.ac.uk).  Juan Ospina is with Los Alamos National Laboratory, Los Alamos, NM, USA ( E-mail: jjospina@lanl.gov).  Charalambos Konstantinou is with the  Computer, Electrical and Mathematical Sciences and Engineering (CEMSE) Division, King Abdullah University of Science and Technology (KAUST), Thuwal 23955-6900, Saudi Arabia (E-mail: charalambos.konstantinou@kaust.edu.edu.)}}
 
}     
\IEEEaftertitletext{\vspace{-2.3\baselineskip}}

\maketitle

\begin{abstract}
The COVID-19 pandemic has impacted our society by forcing shutdowns and shifting the way people interacted worldwide. In relation to the impacts on the electric grid, it created a significant decrease in energy demands across the globe. Recent studies have shown that the low demand conditions caused by COVID-19 lockdowns combined with large renewable generation have resulted in extremely low-inertia grid conditions.  
In this work, we examine how an attacker could exploit these {scenarios} to cause unsafe grid operating conditions by executing load-altering attacks (LAAs) targeted at compromising hundreds of thousands of IoT-connected high-wattage loads 
in low-inertia power systems. 
Our study focuses on analyzing the impact of the COVID-19 mitigation measures on U.S. regional transmission operators (RTOs), formulating a plausible and realistic least-effort LAA targeted at transmission systems with low-inertia conditions, and evaluating the probability of these large-scale LAAs. Theoretical and simulation results are presented based on the WSCC 9-bus {and IEEE 118-bus} test  systems. Results demonstrate how adversaries could provoke major frequency disturbances by targeting vulnerable load buses in low-inertia systems {and offer insights into how the temporal fluctuations of renewable energy sources, considering generation scheduling, impact the grid's vulnerability to LAAs.}
\end{abstract}


\section{Introduction}
\vspace{-1mm}

The cybersecurity of power grids has received significant attention over the past decade{s}. This is mostly attributed to the increase in cybersecurity and cyber-espionage incidents revolving around critical grid infrastructure and energy  delivery systems. Such examples include a large-scale espionage campaign on energy control systems suppliers \cite{cyberthreat1} and 
malware (e.g., Triton) probing the networks of electric utilities \cite{cyberthreat3}. In terms of ongoing research activities, a large body of work is dedicated to investigating the security of bulk power systems  \cite{LakshGT2021, 9261465}. At the same time, the increasing number of Internet-of-Things (IoT) high-wattage consumer appliances including electric vehicles and heating, ventilation, and air conditioning (HVAC) systems, along with the information and communication technologies (ICT) reliance on modern power systems can pose a severe vulnerability to electric grid's operations. The focus of this work is on the less-explored cyber-attacks that target end-user electrical appliances \cite{HamedLAA2011}. 

{An abrupt manipulation of power grid demand by large-scale Botnet\footnote{
{Etymology: The word ``botnet'' is a portmanteau of the words ``(ro)bot'' and ``net(work)'', defined as networks of hijacked computer devices used to carry out various scams and cyber-attacks.}}-type attacks against IoT-smart-home appliances can severely affect the balance between the power supply and demand, and lead to unsafe operation of the grid. Such load-altering attacks (LAAs) that target end-user consumers by targeting their IoT high-wattage devices (e.g,  heater, oven, dryer, etc.) have been examined in literature with results demonstrating that such attacks can lead to high operational costs (at the grid side), unsafe frequency excursions, and even severe frequency and voltage stability issues that can further cause generator trips and cascading failures \cite{dabrowski2017grid, soltan2018blackiot, LakshIoT2021}.}

Moreover, it was shown in \cite{AminiLAA2018} that if attackers manipulate the load over multiple time periods by monitoring the fluctuations of the grid frequencies, they can potentially destabilize the frequency control loop. The information required to execute such attacks can be gathered by publicly available information, such as the charging patterns of plug-in-electric vehicles (PEVs) and the information on the power grid infrastructure \cite{Acharya2020, keliris2019open}. An analytical method to understand the impact of LAAs using the theory of second-order dynamical systems was presented in \cite{LakshIoT2021}. In \cite{9308900}, the authors investigated the impacts of LAAs during low loading conditions. Subsequent work investigated techniques to enhance the resilience of power grids to LAAs by improving the security features for a fraction of smart loads \cite{AminiLAA2018, LakshIoT2021}, and via the use of energy storage to compensate for the destabilizing effect of LAAs \cite{DLAAStorage2020}. Recent work has also investigated data-driven methods to detect and localise LAAs in real time \cite{lakshminarayana2021datadriven}.

{Since the first quarter of 2020}, several researchers have focused their research efforts on analyzing and studying the effects caused by COVID-19 in the operations of electric power systems around the world. 
{For instance, the studies presented in \cite{ieacovid19} and \cite{27europe} show how the electricity demand decreased in different EU countries such as Italy (29\%) and Spain (13.5\%).} In the African region, researchers have investigated the impact of the lockdown measures in the South African power grid where a 20.2\% peak demand reduction and a 28.1\% energy consumption reduction were observed \cite{28africa}. Moreover, other regions in the world, such as the East Asian region, have seen significant electricity consumption reductions with around 13\% demand drops \cite{energiesCOVID}. Comprehensive reviews of the mid-to-long range impacts of the pandemic in the energy sector are presented in \cite{energiesCOVID, ieacovid19, appliedEnergy}. 
{In these studies, the effects of COVID-19 are analyzed and reviewed for different regions in the world, 
diverse types of power systems (e.g., transmission, distribution, renewables, etc.), and a variety of energy operations (e.g., load/renewable energy forecasting, energy markets, peak demand, etc.). A common conclusion of this research is that the COVID-19 pandemic and the lockdown measures related to it, especially in the first half of 2020, resulted in low loading conditions with higher renewable energy penetration in power grid infrastructures. }


In this work, 
{we further investigate how transmission systems with high penetration of renewable energy resources (RES), and subsequently the supporting low-inertia conditions during the initial outbreak of COVID-19, could allow attackers to leverage the system conditions towards the realization of LAAs.} {The analysis is particularly important, given the significant increase in the number of {cybersecurity-related} incidents since the COVID lockdown measures were announced \cite{LALLIE2021102248}.}
We analyze generation and load data from Regional Transmission Operators (RTOs) in the U.S. and examine how the system inertia in areas with high penetration of intermittent RES (such as PV and wind energy) is impacted by transients and small-signal stability issues causing them to be more vulnerable to LAAs. Different from \cite{9308900}, this work focuses on analyzing how plausible is the implementation of high-impact LAAs in a low-inertia power system with high penetration of RES. The impacts are analyzed based on the RES penetration of the system. The least-effort location and amount of load to compromise the system are identified based on a novel LAA formulation targeted at low-inertia power networks.

Furthermore, we present a theoretical framework to assess how the low-inertia conditions due to the high penetration of RESs adversely affects the system's resilience to LAAs. Our analysis is based on the theory of second-order dynamical systems applied to the power grid model \cite{LakshIoT2021}. Specifically, we derive the sensitivity of the power grid's dynamical response (to LAAs) with respect to the change in the system inertia at different generator nodes. Using these results, we quantify the magnitude of LAAs that can cause unsafe frequency excursions in light of the low-load high RES penetration conditions caused due to COVID-19. {Furthermore, we shed light on how the uncertainty of renewable energy generation and generator scheduling impact the grid's vulnerability to LAAs.}

{The contribution and novelty of our work can be summarized as follows:}
\vspace{-1mm}
\begin{itemize}[itemsep=0pt,parsep=0pt,leftmargin=*, wide=0pt]
    \item 
    {Despite the literature on IoT-based LAAs \cite{dabrowski2017grid, soltan2018blackiot, AminiLAA2018, LakshIoT2021, Acharya2020, 9308900}, none of these works have considered the low-inertia conditions caused due to growing RES penetration and a pandemic-type event. The COVID-19 energy data offers a unique insight into these conditions, which we utilize to illustrate our results. }
    \item 
    {While COVID-19 {related} data has been analyzed in different contexts \cite{ieacovid19, 27europe, EUPrince2020,BADESA2021}, none of these works considered the analysis from a cybersecurity perspective. Our work offers novel insights into how the temporal fluctuations of RES, considering generation scheduling, impact the grid's vulnerabilities to LAAs.}
    \item 
    {Prior work has only analyzed LAAs by performing exhaustive simulations \cite{dabrowski2017grid, soltan2018blackiot, AminiLAA2018}. However, such an analysis would be computationally expensive, since the operator must perform simulations under several inertia conditions and several combinations of nodes that could be subject to attack{s}. To overcome this drawback, we derive theoretical expressions for the sensitivity of the system's eigensolutions with respect to the inertia of the system using the theory of second-order dynamical systems, which in turn can be used to predict the change in the system's response due to RES penetration and the magnitude of LAAs that leads to unsafe events. }
\end{itemize}

{The rest of the paper is structured as follows. In Section II, we present the threat model for the proposed LAA and a realistic scenario on how such LAAs can be realized with existing vulnerabilities on HVAC systems.} Section III presents an analysis of how COVID-19 lockdown measures affected load conditions and operations in power systems. Section IV demonstrates the theoretical analysis to characterize the effect of low load and reduced inertia on the power grid dynamics under LAAs, while Section V presents the experimental setup and results. Finally, Section VI concludes the paper and provides directions for future work. 

\vspace{-2mm}
\section{Load-Altering Attack Scenario on IoT Controllable Loads}
\vspace{-1mm}
In this section, we demonstrate the details of an example attack scenario from commercial IoT-controllable high wattage loads,  
and how such an attack can maliciously affect the stability of the grid, causing degradation of grid equipment or even power outages and large-scale blackouts. Specifically, we provide the details of attack exploitation of 
an HVAC system (load-side) which can lead to tampering of operation information and system configuration, leading even to denial-of-service (DoS) conditions. 
A similar type of attack can also be performed on the {distributed energy resources} (DER) side, e.g., solar inverters, in which firmware backdoors could enable access to weakly encrypted user passwords, which could then be reversed allowing unauthorized access \cite{solarinverter}, {\cite{9430606}}. 

\vspace{-2mm}
\subsection{Air-Conditioner Load-Altering Attack}

The 
air-condition{ing}
(AC) industry market {reached} a 100 billion USD annual revenue {in 2021}~\cite{9096505}, primarily due to a steady raising in electricity consumption and peak demand for electricity. 
For example, in India, the usage of electricity attributed to ACs will reach 239 TWh/year by 2030 which will require an additional 500 MW generation capacity \cite{SINGH2018432}. ACs role, particularly for the peak load during long and hot summers, is of paramount importance. For instance, in California, U.S. just the commercial AC usage constitutes almost 45\% of the peak power demand \cite{de2019global}. Furthermore, the growing prevalence of ACs, considering that 20\% of the global electricity demand is consumed by cooling loads \cite{alkhraijah2021effects}, as well as the load demand increase during the initial spread of COVID-19 \cite{9308900}, requires power utilities to handle challenges related to energy imbalance, especially with the 
incorporation of RES \cite{8219356}.

\subsubsection{Vulnerability Description} Mitsubishi Electric has recently announced that a large range of their AC systems have been identified with a number of vulnerabilities able to cause information disclosure, privilege escalation, and DoS. Specifically, the centralized controllers of those systems were found to be vulnerable due to ``improper restriction of XML external entity reference (XXE)'' (CWE-611)\footnote{\url{https://cwe.mitre.org/data/definitions/611.html}}, i.e., the affected products do not restrict XML external entity references in a sufficient manner, and thus, redirect software processes outside of the intended sphere of control \cite{mitsibushi1}. This vulnerability can be triggered by sending XXE payload to the process listening to the TCP port number 1025, which causes the application to make arbitrary HTTP and/or FTP requests. As a result, this vulnerability (CVE-2021-20595), if exploited, can allow malicious adversaries to disclose AC operational settings and data as well as access arbitrary files on the system, or even cause DoS attacks by forming and transmitting specially crafted packets.

The same models of Mitsubishi Electric AC systems are also vulnerable to privilege escalation weaknesses resulting from improper implementation of authentication algorithms \cite{mitsibushi2}. Such vulnerabilities of improper authentication (CWE-303)\footnote{\url{https://cwe.mitre.org/data/definitions/303.html}}, may allow malicious adversaries to perform impersonation attacks, and therefore, manipulation operating data and configuration settings of the AC systems (CVE-2021-20593). 

\subsubsection{Attack Scenario and Impact}
Both described vulnerabilities can be exploited remotely{. Thus,} in our attack scenario, we consider an oblivious adversary with limited or even zero knowledge about the details of the grid system including system topology, lines parameters, and component interconnection details, who cannot physically manipulate the AC system asset under attack, and has sufficient resources that allow him/her to have access to the network control of the AC \cite{zografopoulos2021cyber}. The limited adversary knowledge is realistic due to the  restricted access to cyber-physical system control and errors in the data collection process.

\begin{figure}[t]
    \centering
    \includegraphics[width=2.5in]{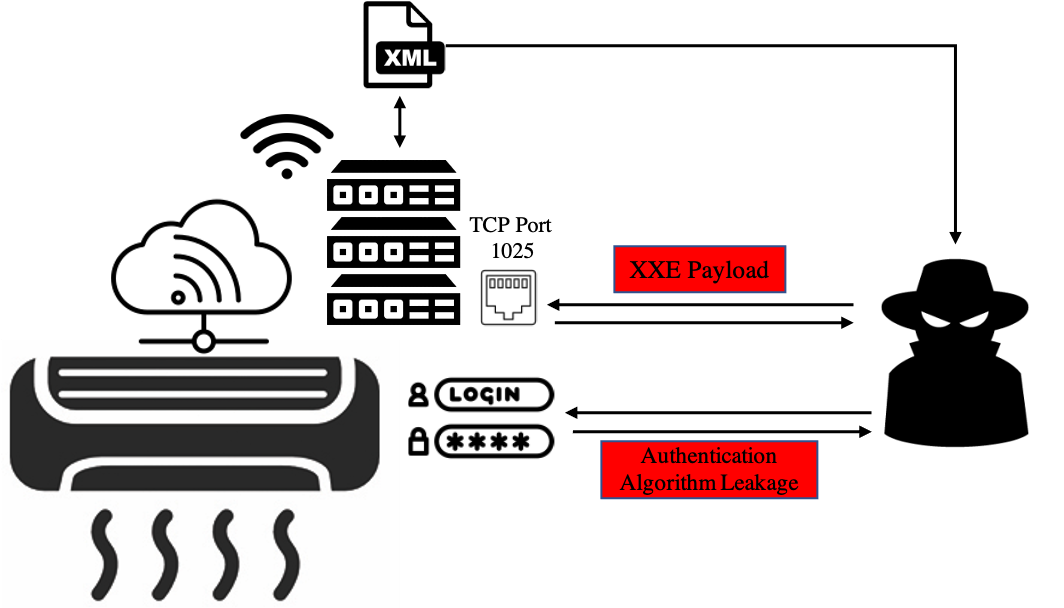}
    \vspace{-1mm}
    \caption{Graphical depiction of vulnerability exploitation towards AC load-altering attack {(LAA)}.}
    \vspace{-1mm}
    \label{fig:attackscenario}
\end{figure}

In the attack scenario, as shown in Fig. \ref{fig:attackscenario}, we consider the adversary able to trigger the CWE-611 vulnerability by sending XXE payloads to the process listening to the TCP port 1025. Such a payload can make the application process an XML document that includes XML entries with URIs (uniform resource identifiers). Those URIs being resolved to arbitrary HTTP and/or FTP requests beyond the intended sphere of control, lead to the IoT-controllable AC embedding incorrect documents into its output, i.e., due to improper restriction of XML XXE, the AC application may echo back the data (e.g., in an error message), thereby exposing the file contents. {The} data could include operational settings of the device such as nominal load conditions, historical settings and usage, modes of operation, and even authentication data. The leakage of authentication data, such as weakly encrypted user passwords due to the described vulnerability issue CWE-303 \cite{keliris2017ge}, could allow adversaries to reverse them enabling unauthorized access. An attacker able to remotely access such units on a large scale could drive the realization of LAAs.

\begin{figure}[t]
    \centering
    \includegraphics[width=3.1in]{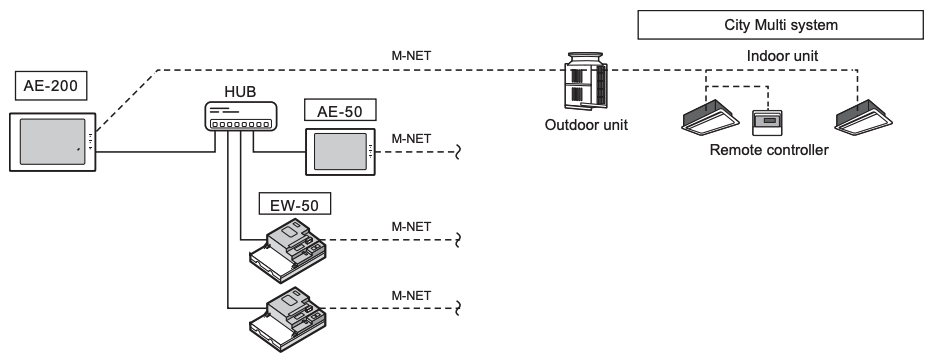}
    \vspace{-1mm}
    \caption{EW-50A Mitsubishi controller configuration when controlling more than 50 units of equipment \cite{controllerEW50A}.}
    \vspace{-1mm}
    \label{fig:controllersetup}
\end{figure}

To further enhance the realism of such an attack scenario, consider the EW-50A controller, an affected product of the Mitsubishi Electric family \cite{controllerEW50A, mitsibushi1}. This AC controller can operate either as a stand-alone central controller or network up to three EW-50A expansion controllers with one AE-200A controller. The AE-200A controller can manage up to 50 indoor units individually or up to 200 indoor units when installed with three AE-50A 
controllers \cite{controllerAE200A}. Fig. \ref{fig:controllersetup} shows a typical 
configuration with more than 50 units of ACs being able to connect via the AE-200 and EW-50 
setup. In a typical scenario, a building configuration would include 50 ceiling cassette AC units, each one able to cool an area of up to 147m$^2$. Such air cooling (and purifying) cassettes have a typical kW range of operation between 7.1 -- 15.8 kW. 
Therefore, an AC LAA following the aforementioned steps could cause, for instance, 50 units during \textit{eco}-mode of operation in which the AC compressor runs slower consuming the minimum (e.g., 7.1 kW), move towards \textit{turbo}-mode operating at the maximum capacity (e.g., 15.8 kW), essentially doubling the consumption energy per hour. At the 50 unit scale of a commercial building, that is approximately 
435 kW additional power. 

\subsection{Threat Model for AC Load-Altering Attack}

In this subsection, we dive deeper into the specific threat model used to describe the AC LAA. {Table \ref{tab:threatmodel} shows the threat and vulnerability modeling of the LAA discussed. This table is based on the threat model proposed in \cite{zografopoulos2021cyber}. As seen in the table, the threat is characterized based on different characteristics that describe important information related to the accessibility, assets, and techniques (among other characteristics) needed to deploy the LAA.}

{As previously mentioned, for the described LAA, the attacker is considered to be a \textit{semi-oblivious} attacker, which indicates that {she/he} possesses limited information of the system, with a \textit{targeted} specificity (i.e., the attacker specifically targets vulnerabilities in high-wattage loads connected to the system) and \textit{non-possession} accessibility (i.e., the attacker does not need to physically interact or possess the attacked asset). In terms of resources, we can catalog the proposed LAA as a \textit{Class II} adversary, where the adversary needs or has sufficient motivation and resources to carry out the attack without being detected or compromised. A \textit{Class I} adversary would indicate an attacker without sufficient resources or motivation to carry out complex attacks. For frequency and reproducibility, the LAA presented is designed as an \textit{iterative} and \textit{multiple-times} attack, since, in order to cause the corresponding damage, the attack must be iteratively deployed at multiple load buses and must be deployed multiple times so the system is destabilized. Moreover, the LAA threat is considered to have an attack functional level of \textit{L1}, where the assets (e.g., smart HVAC and high-wattage IoT-connected devices) are compromised by \textit{control manipulation and/or modification} while existing in the industrial network. Finally, the premise of the LAA is categorized as \textit{Cyber: Integrity} due to the fact that the threat can be considered to belong to a subset of data integrity attacks (DIA), where the integrity of the control system is compromised.}

\begin{table}[]
\centering
\caption{Threat model formulation of AC LAA.}
\label{tab:threatmodel}
\begin{tabular}{||c|c||}
\hline \hline
\textbf{Threat Model \textbackslash Threat} & LAA \\ \hline
\textbf{Knowledge} & \begin{tabular}[c]{@{}c@{}}Semi-Oblivious\end{tabular} \\ \hline
\textbf{Access} & Non-possession \\ \hline
\textbf{Specificity} & Targeted \\ \hline
\textbf{Resources} & Class II \\ \hline
\textbf{Frequency} & Iterative \\ \hline
\textbf{Reproducibility} & Multiple-times \\ \hline
\textbf{Attack Func.\ Level} & L1 \\ \hline
\textbf{Asset} & \multicolumn{1}{l||}{\begin{tabular}[c]{@{}l@{}}~~~~~~~Smart HVAC, \\High-wattage IoT devices\end{tabular}} \\ \hline
\textbf{Technique} & \multicolumn{1}{l||}{\begin{tabular}[c]{@{}l@{}}Control modification\end{tabular}} \\ \hline
\textbf{Premise} & Cyber: Integrity \\ \hline \hline
\end{tabular}
\end{table}


\vspace{-2mm}
\section{Analysis of COVID-19 lockdowns in Electric Power Systems}
{During the initial outbreak of COVID-19 in early 2020, the electricity consumption of our society had a dramatic shift since most business, manufacturing, and urban centers were forced to close down. Lockdown measures and state-at-home orders (SAHOs) were issued in major cities around the U.S. and the world, shifting the spatial and temporal energy load consumption and causing major net-load reductions throughout energy systems. The low loading conditions coupled with high renewable energy penetration may pose a threat to power systems due to the inherent low inertia of these systems. In this work, we analyze the impact of the lockdown measures to determine if and how an attacker can exploit these conditions and cause harm to electric power systems. In particular, we focus on large-scale IoT-enabled LAAs against the power grid, in which the attacker causes an abrupt shift in the energy demand, that can potentially cause unsafe frequency fluctuations and/or destabilize power grid control loops.}

\begin{figure*} \centering  
\centerline{\includegraphics[width=0.75\textwidth]{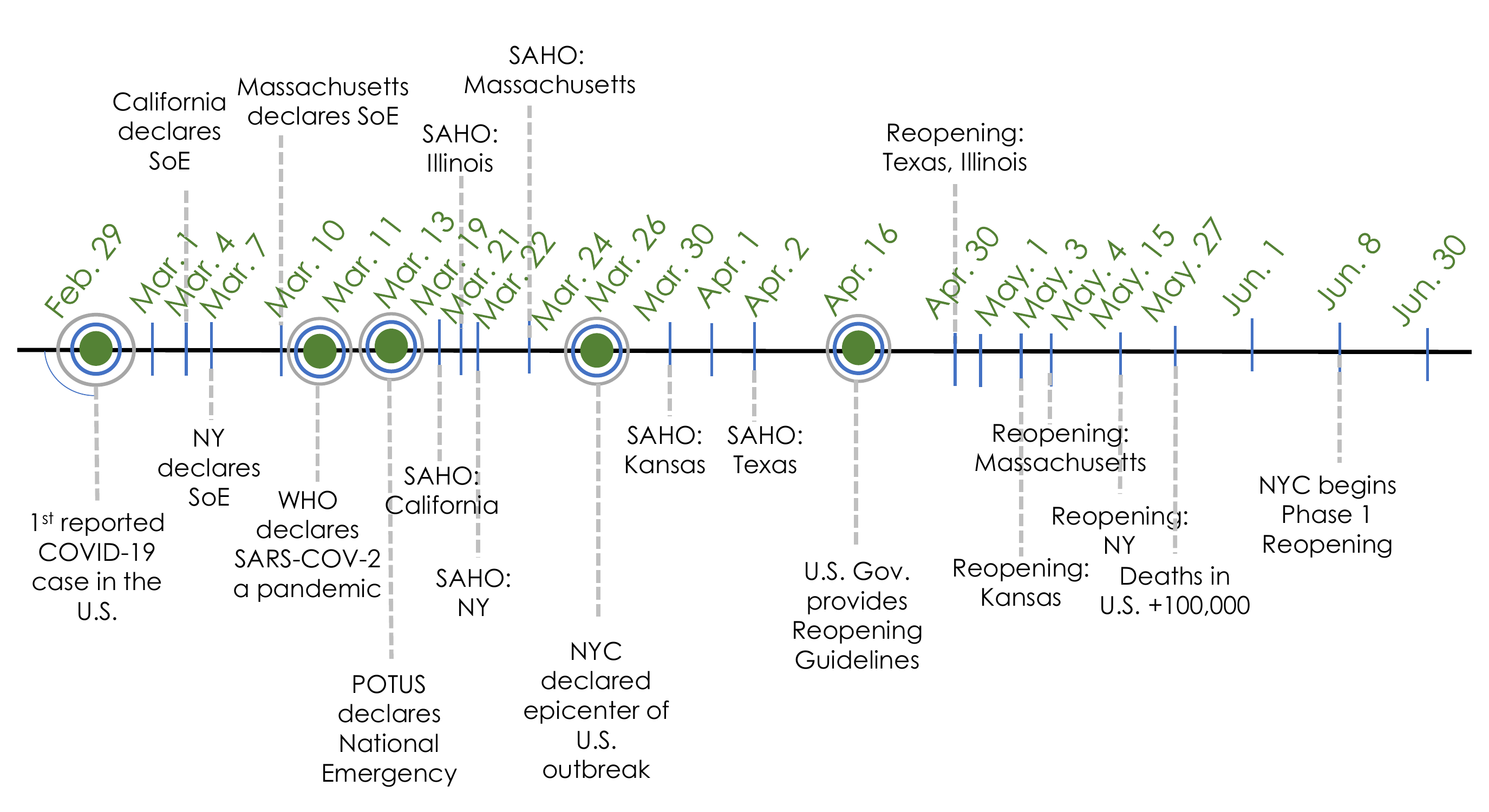}}
\vspace{-1mm}
\caption{COVID-19 lockdown-measures timeline in the U.S. for the year 2020.}
\label{fig:resp_timeline}
\vspace{-0.4 cm}
\end{figure*}

\subsection{Analysis of Lockdown Measures Effects in 
Power Systems}

For the proposed analysis, we analyze the data obtained from the cross-domain open-source data hub COVID-EMDA \cite{ruan2020cross}, focusing on examining the RES penetration, in terms of total load percentage, and load reduction during the lockdown measures of 2020 of seven different U.S. RTOs: The Electric Reliability Council of Texas (ERCOT), Independent System Operator-New England (ISO-NE), Midcontinent Independent System Operator (MISO), New York Independent System Operator (NYISO), Pennsylvania-New Jersey-Maryland (PJM) Interconnection, California Independent System Operator (CAISO), and Southwest Power Pool (SPP). 

In order to identify the most vulnerable electric grid region due to low loading conditions caused by COVID-19, it is necessary to examine the lockdown response timeline of different states throughout the U.S. We compile a series of important events, related to the COVID-19 outbreak from different sources such as \cite{covidrestrict1, covidrestrict2}, with the objective of finding a correlation between the load variation and these events. Fig. \ref{fig:resp_timeline} shows the COVID-19 lockdown response timeline for different places inside the U.S. RTOs analyzed during the 2020 COVID-19 pandemic. As seen in this timeline, most of the SAHOs and lockdown measures were concentrated during the period ranging from March 1st to June 30th, 2020, while most of the re-openings at major cities occurred during the second and third weeks of the month of May. 

\begin{figure*}[h] \centering 
\subfigure[] { \label{fig:5}     
\includegraphics[width=5.72cm]{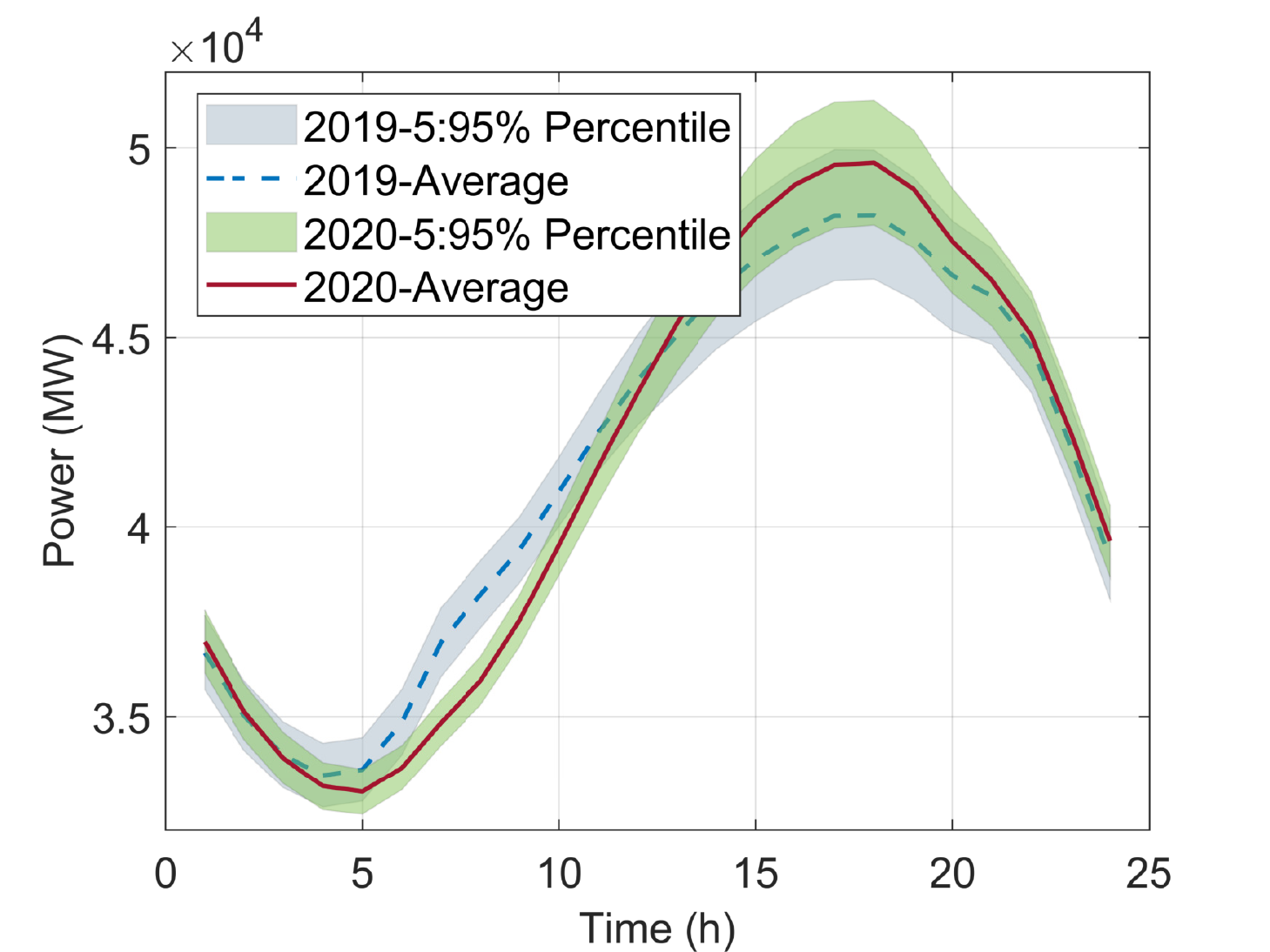}}
\subfigure[] { \label{fig:6}     
\includegraphics[width=5.7cm]{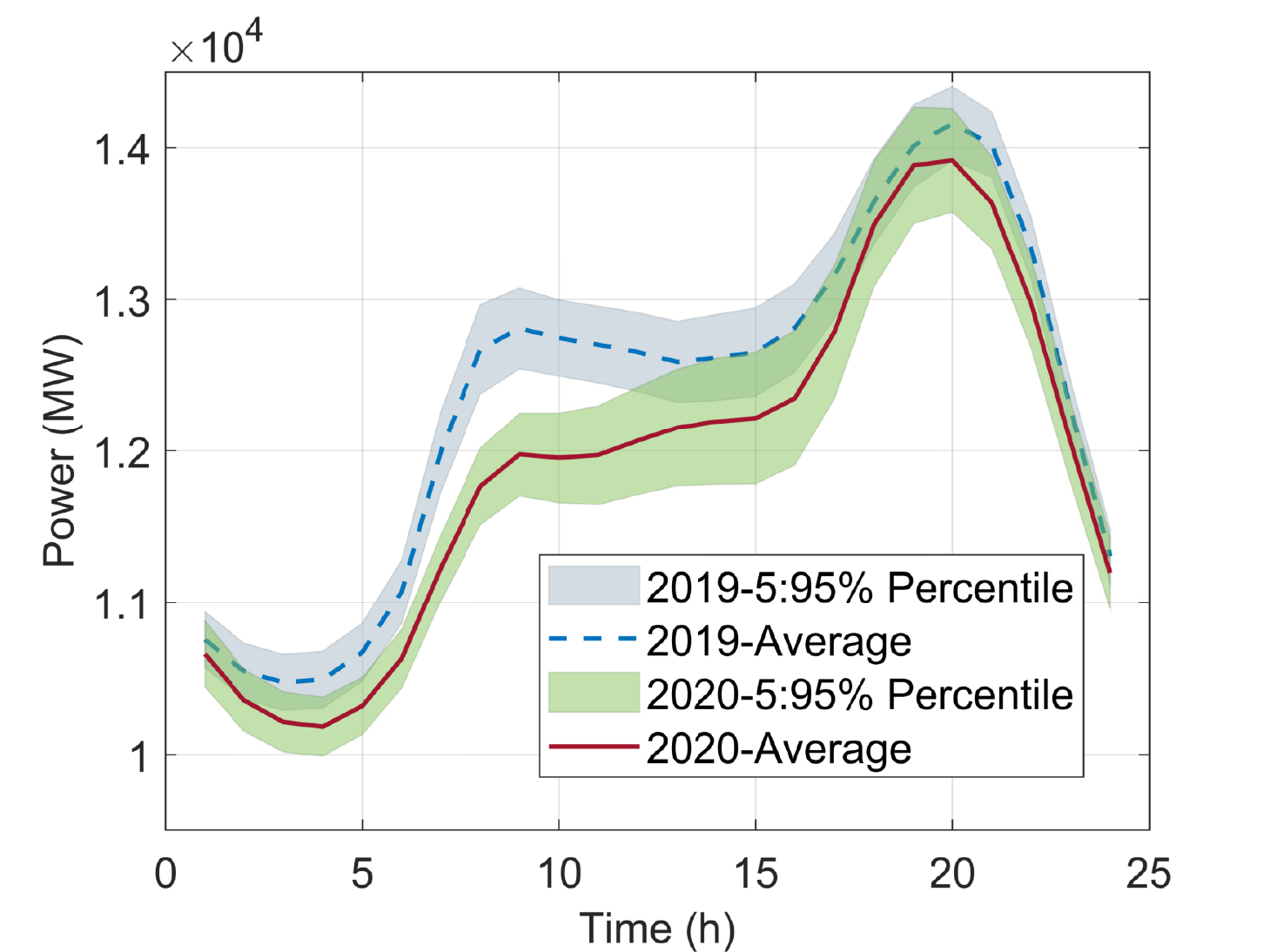}}
\subfigure[] { \label{fig:7}     
\includegraphics[width=5.7cm]{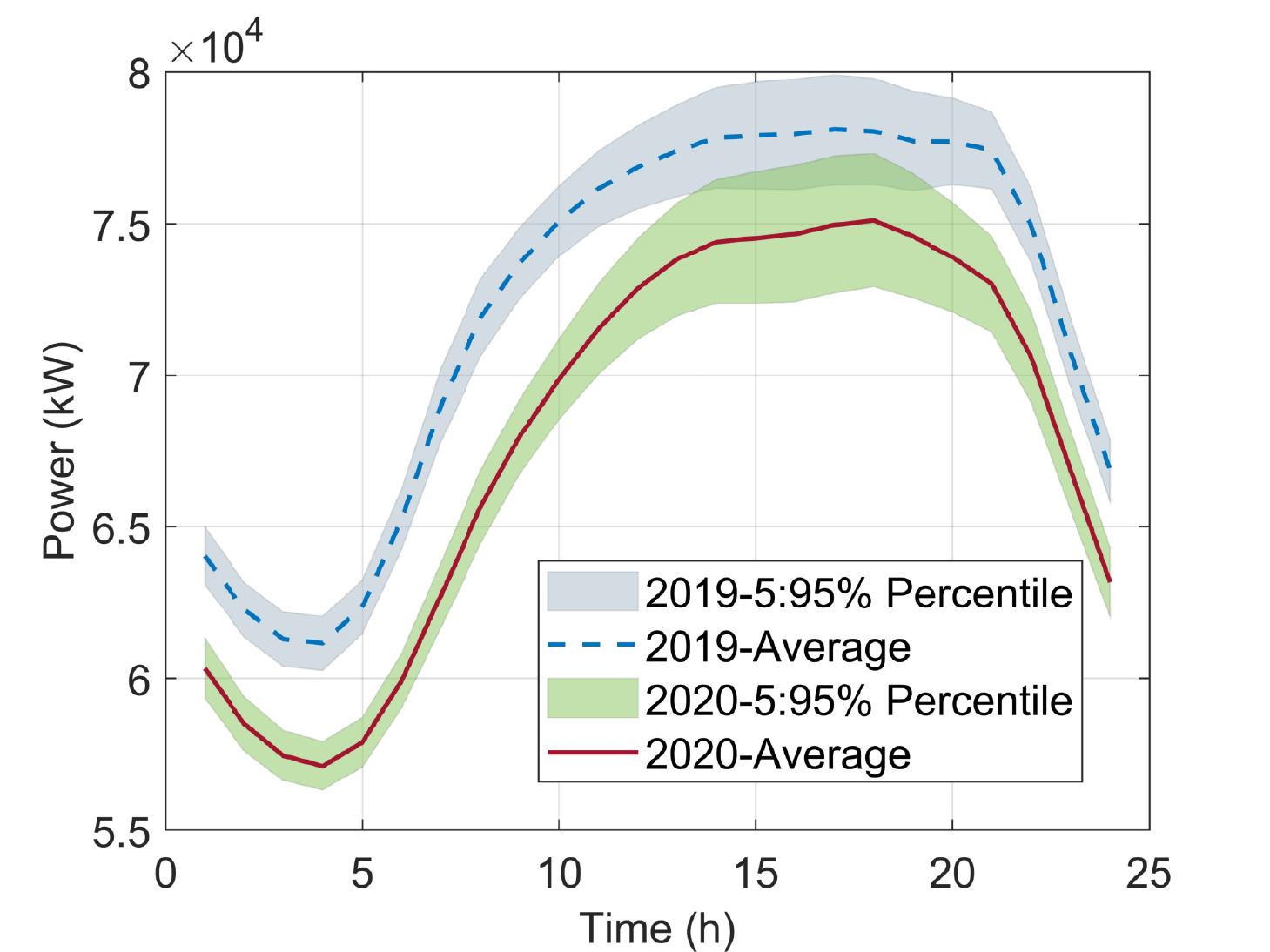}}
\\
\subfigure[] { \label{fig:8}     
\includegraphics[width=5.7cm]{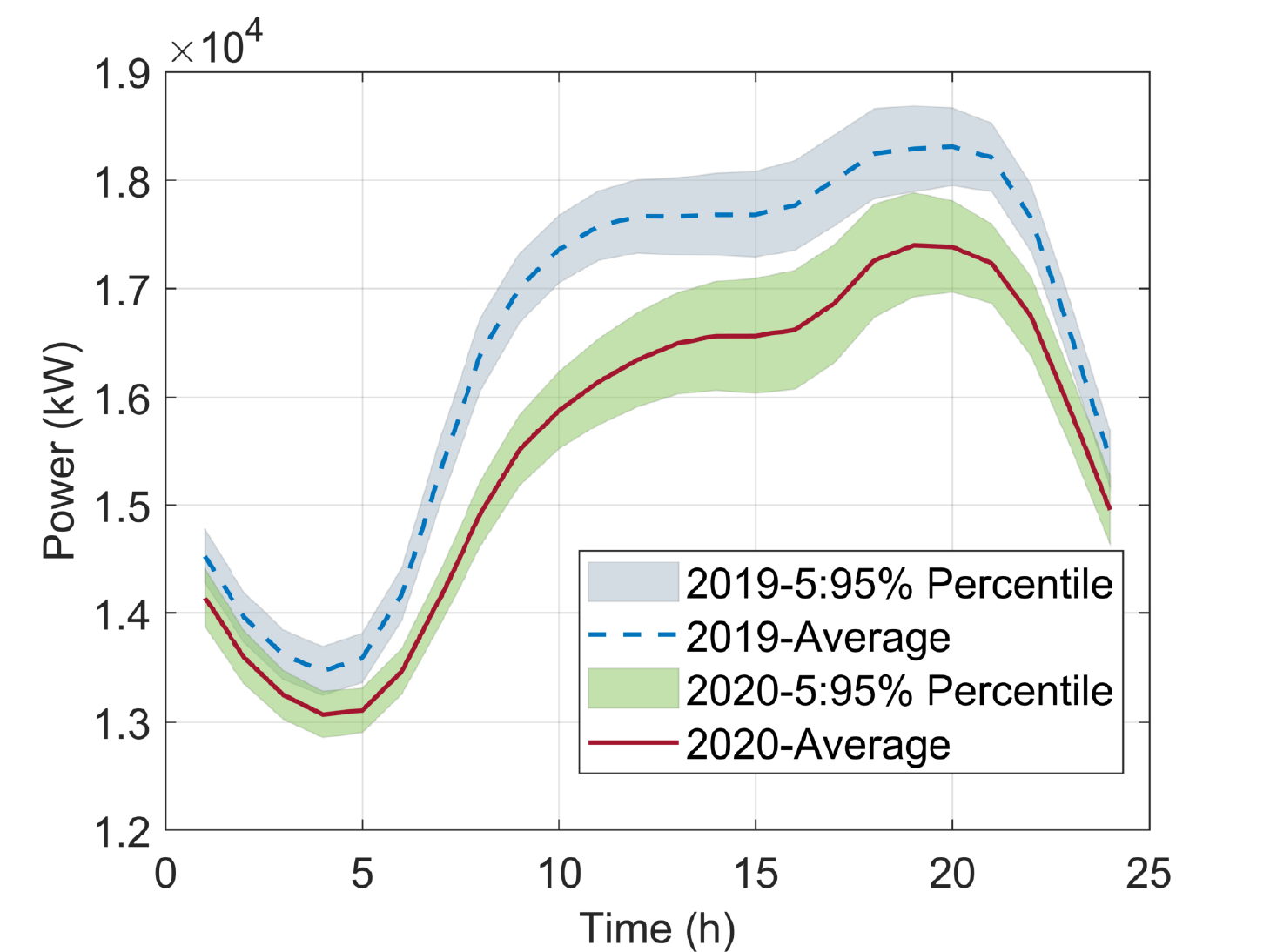}}
\subfigure[] { \label{fig:10}     
\includegraphics[width=5.7cm]{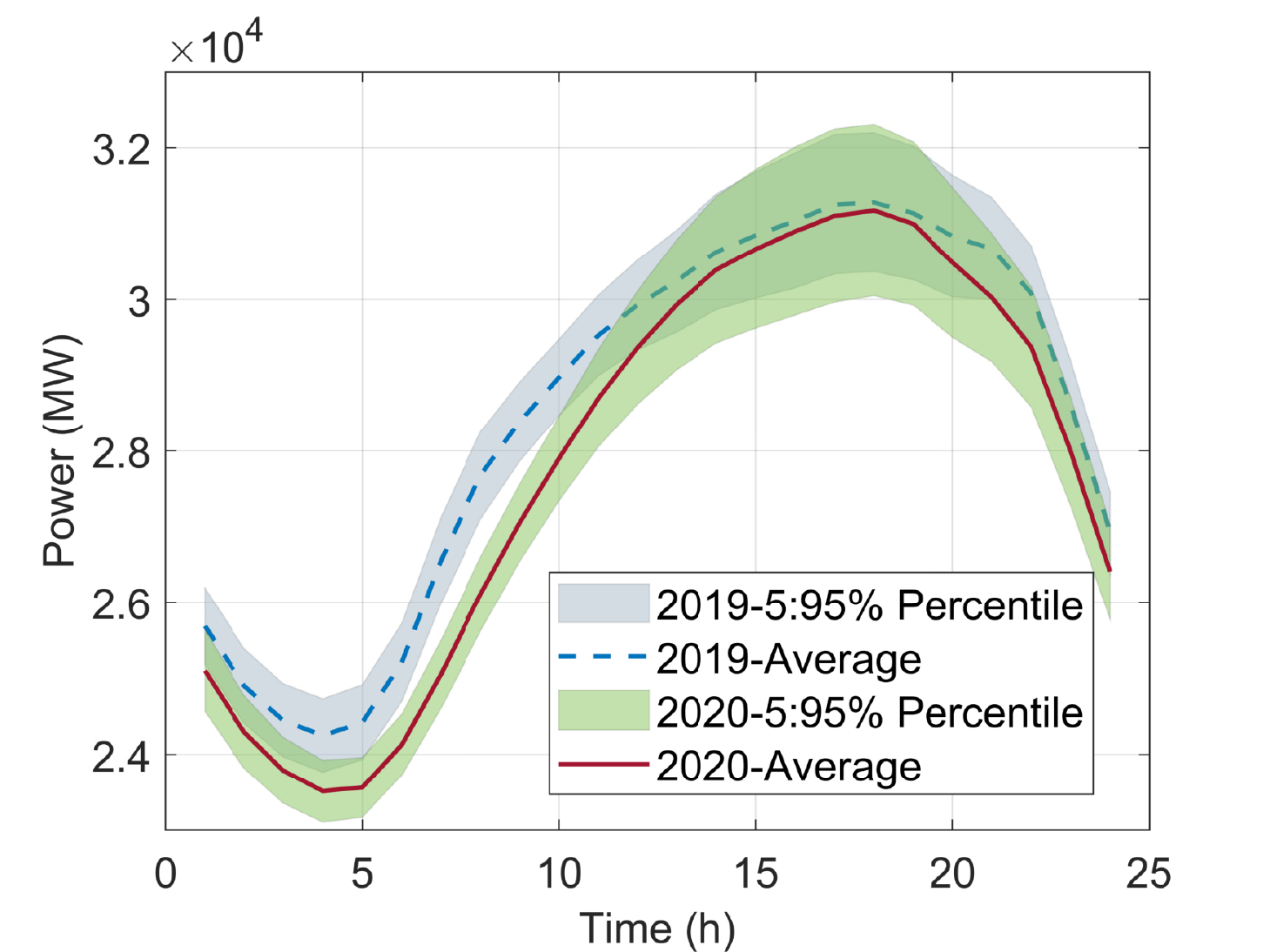} }
\subfigure[] { \label{fig:11}     
\includegraphics[width=5.7cm]{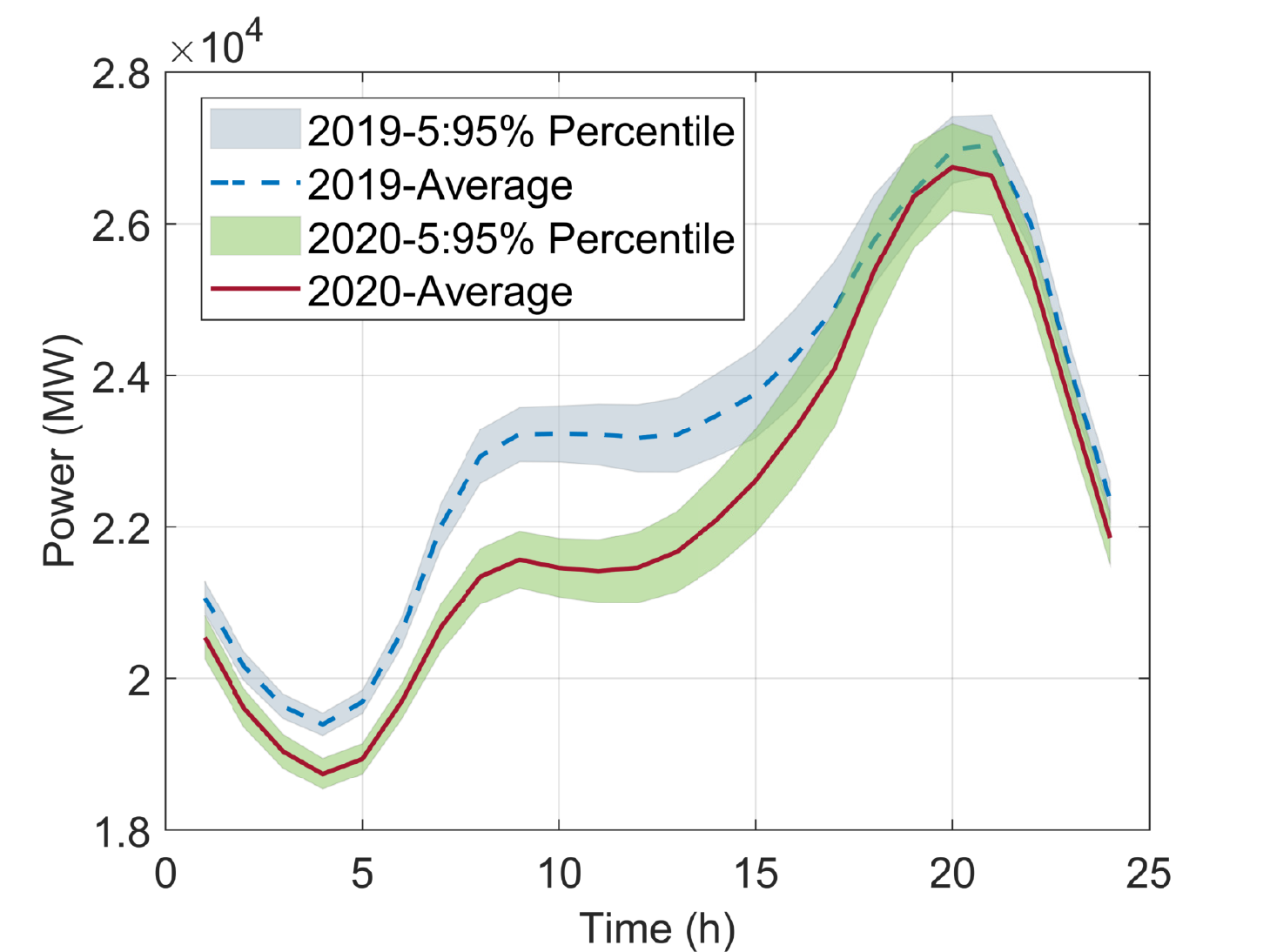} }

\caption{2019 vs. 2020 total load in every U.S. RTO, 5\% to 95\% percentiles for: a) ERCOT, b) ISO-NE, c) MISO, d) NYISO, e) SPP, and f) CAISO. \textit{(PJM is also analyzed but excluded due to space limitation)}.} 
\label{fig:loadscomparison}  
\end{figure*}

Based on this timeline, we combine the load consumption and RES generation data from the different RTOs and analyze them to determine the load reduction and RES penetration of different U.S. RTOs  during COVID-19. For the first analysis, we process and analyze the load consumption data in order to determine how the lockdown measures affected the load consumption throughout the different RTOs. Fig. \ref{fig:loadscomparison} shows the 5-95\% percentiles of the load demand at the different RTOs for the 2019 and 2020 March 1st-to-June 30th periods. This figure showcases the significant reduction of load consumption at RTOs such as MISO, NYISO, SPP, and CAISO, while RTOs such as ERCOT and ISO-NE are the ones with smaller load reductions. It is important to notice that the major net-load demand reductions throughout all RTOs are observed during the `morning' hours (between 5 am-12 pm). According to this analysis, the RTOs with the highest average load difference between 2019 and 2020 are CAISO, NYISO, and MISO, with a maximum average load difference of 1804 MW at 11 am,  1500 MW at 9 am, and 6278 MW at 7 am, respectively.

\begin{figure*} \centering    
\subfigure[] { \label{fig:12}     
\includegraphics[width=5.72cm]{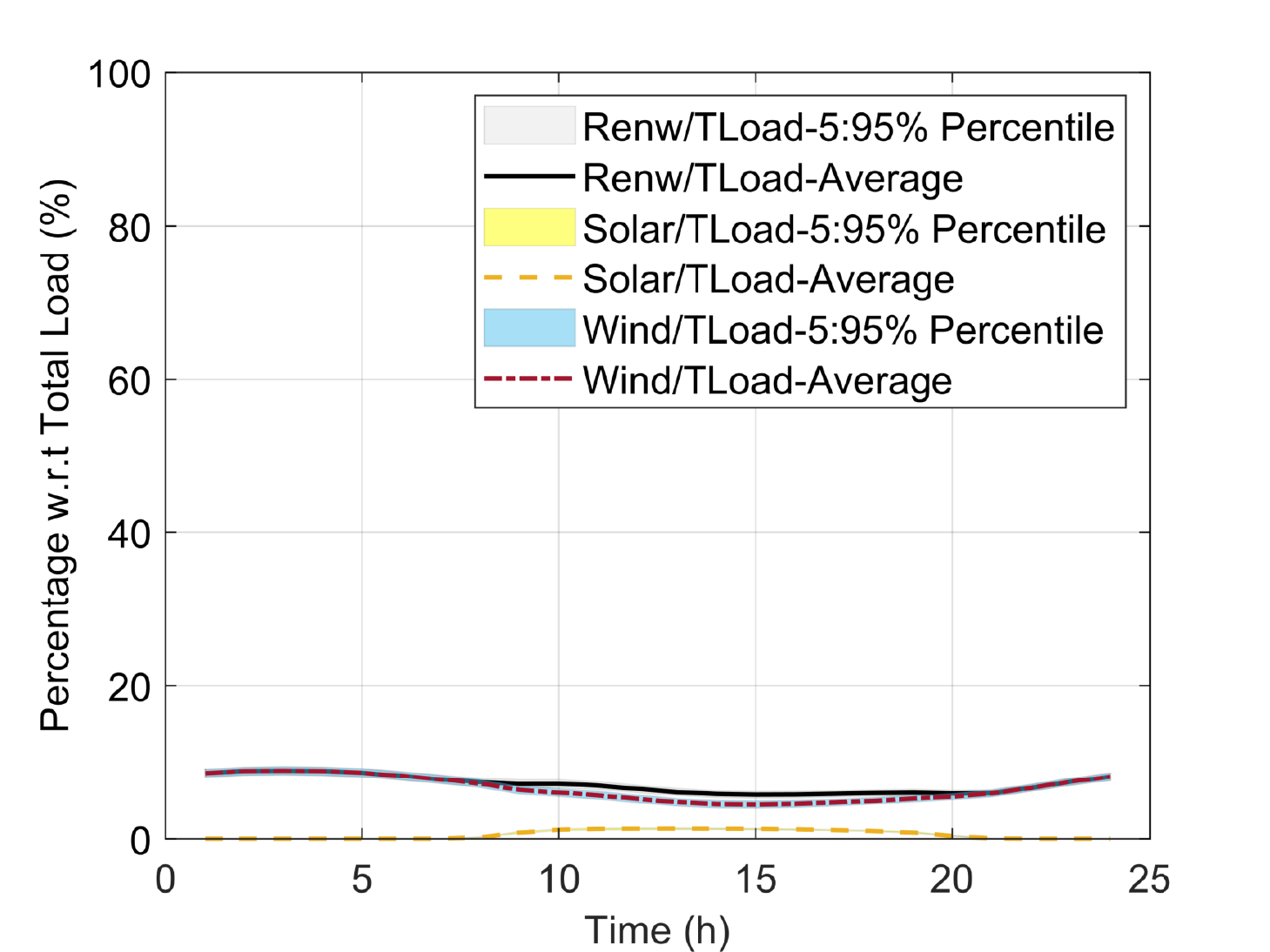}}
\subfigure[] { \label{fig:13}     
\includegraphics[width=5.7cm]{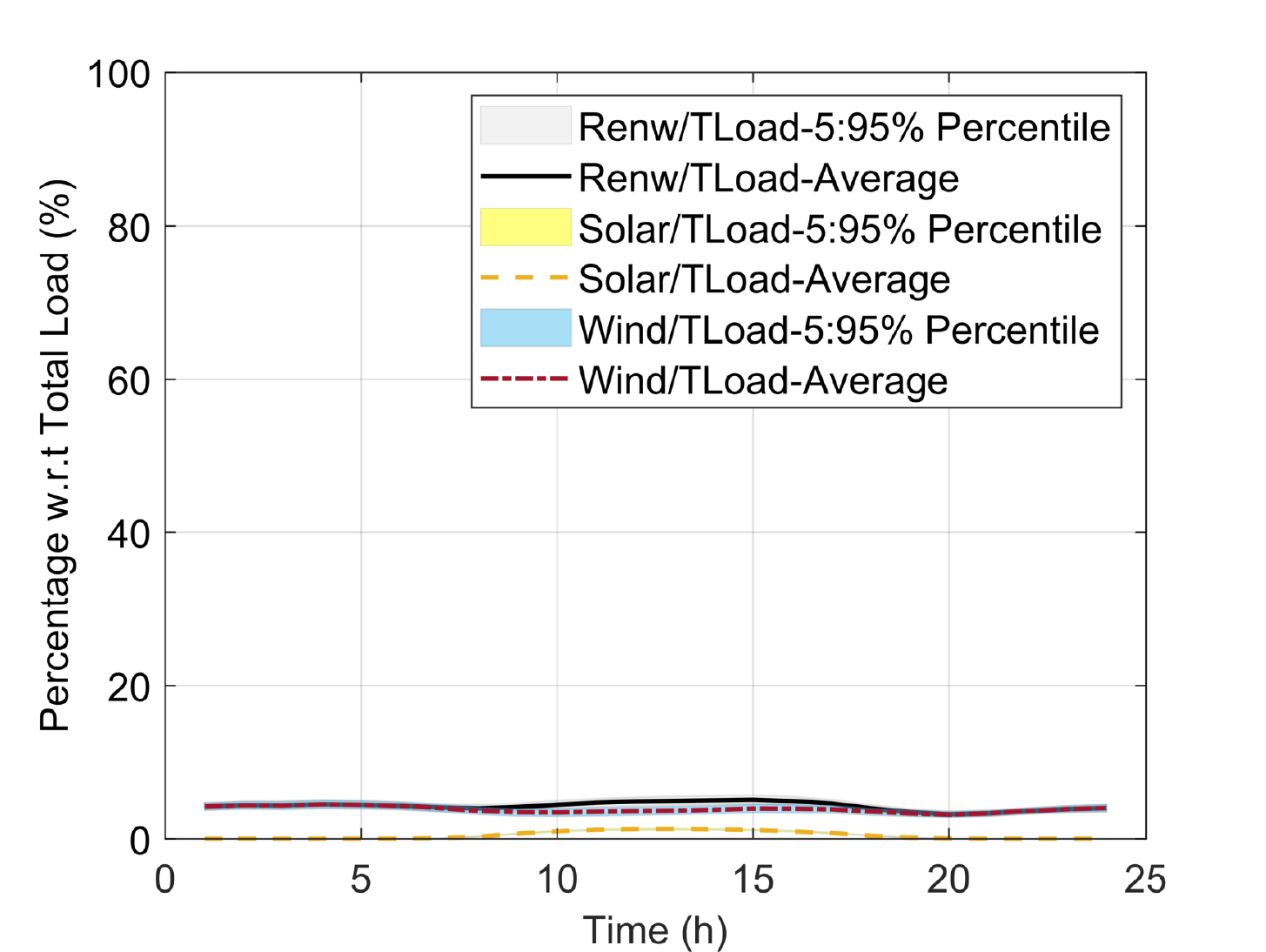}}
\subfigure[] { \label{fig:14}     
\includegraphics[width=5.7cm]{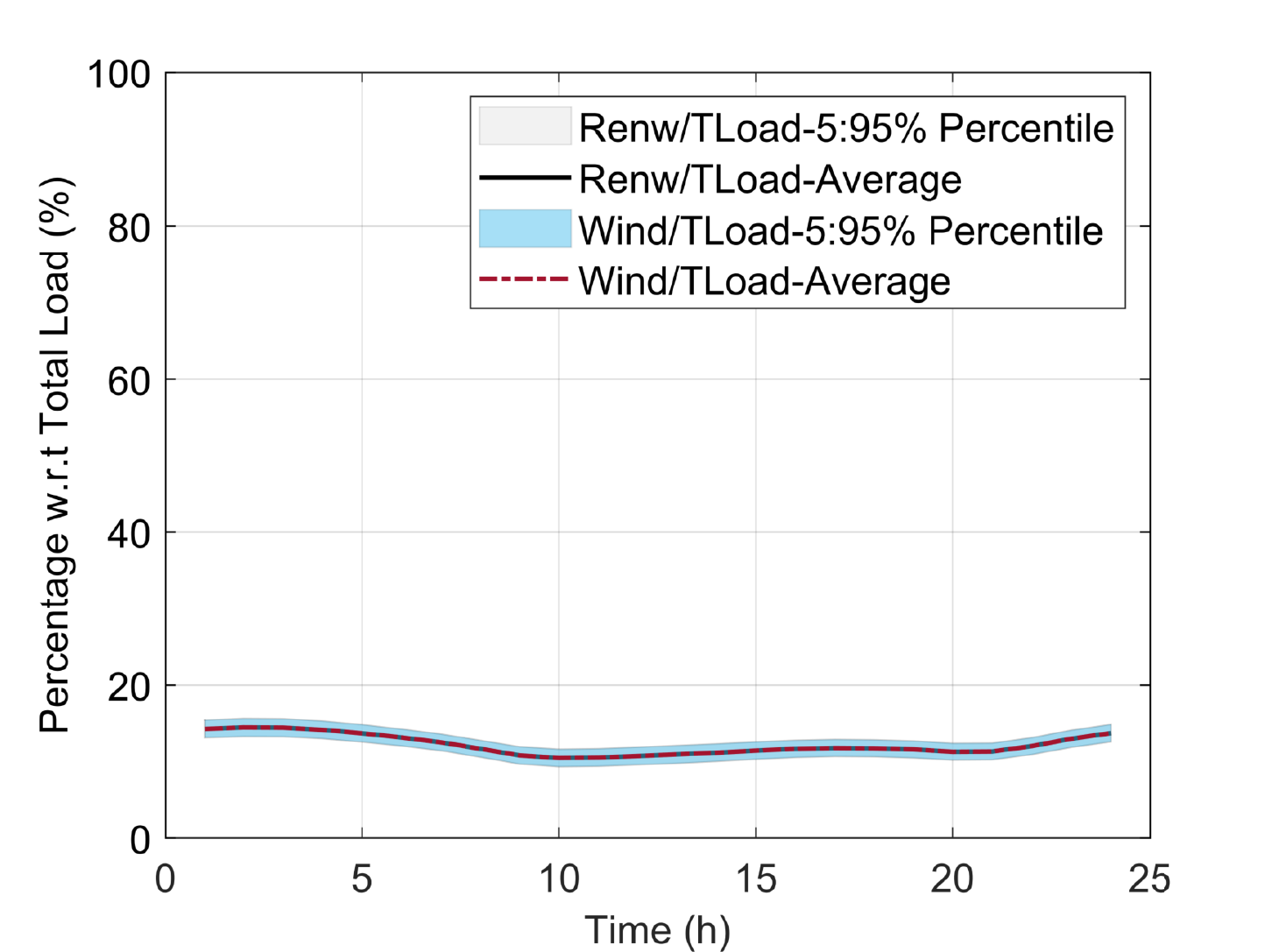}}
\\
\subfigure[] { \label{fig:15}     
\includegraphics[width=5.7cm]{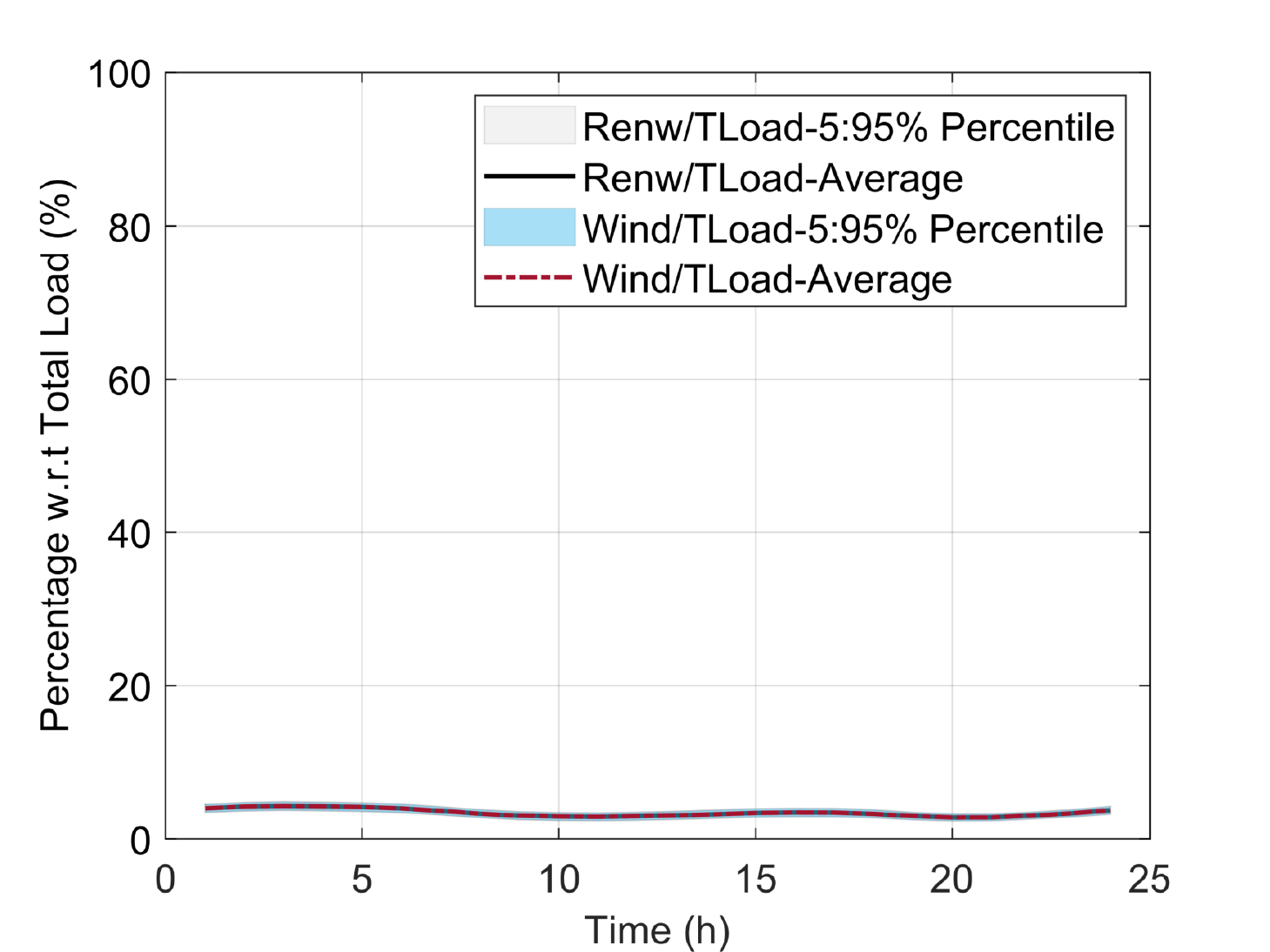}}
\subfigure[] { \label{fig:17}     
\includegraphics[width=5.7cm]{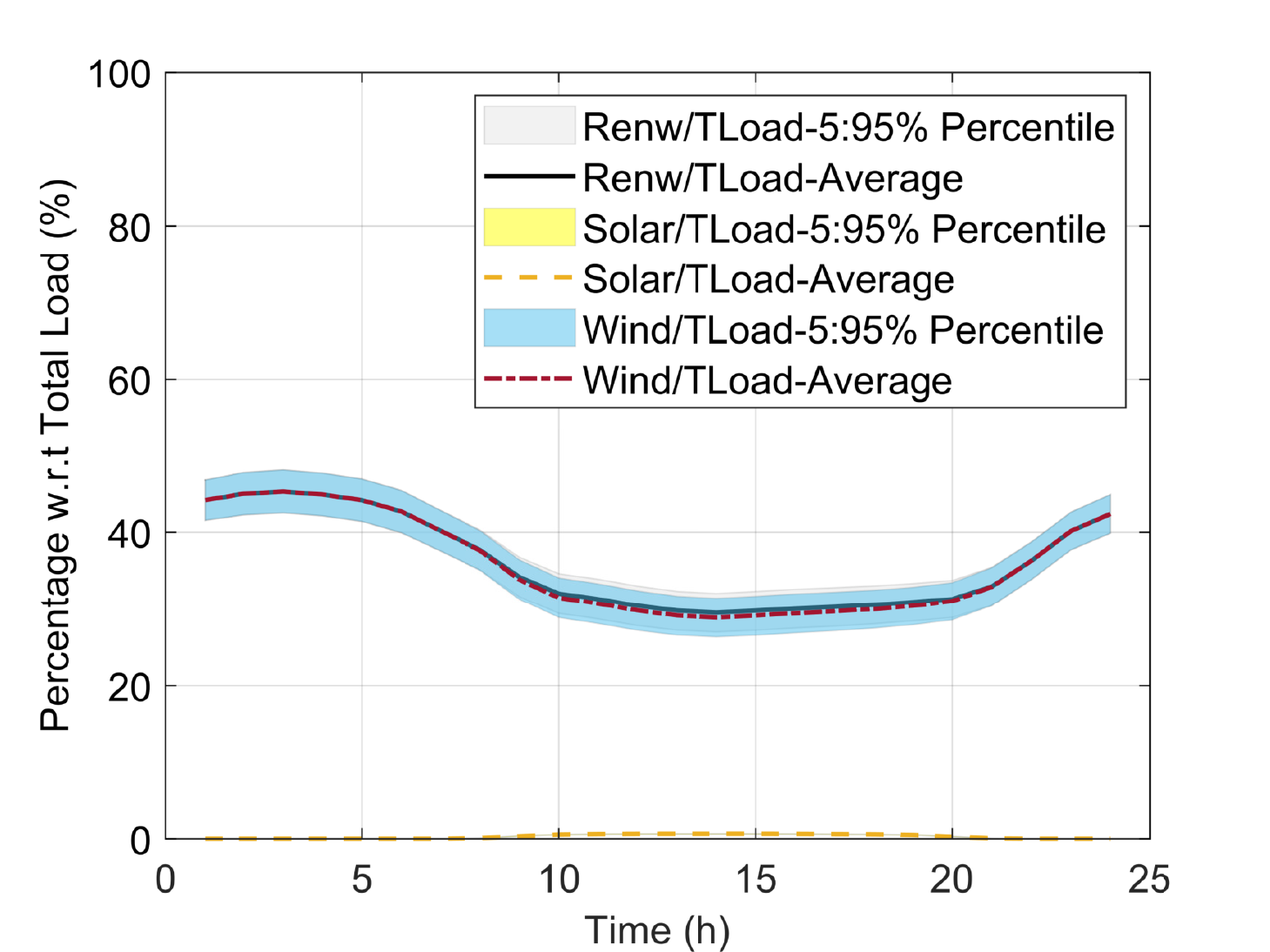} }
\subfigure[] { \label{fig:18}     
\includegraphics[width=5.7cm]{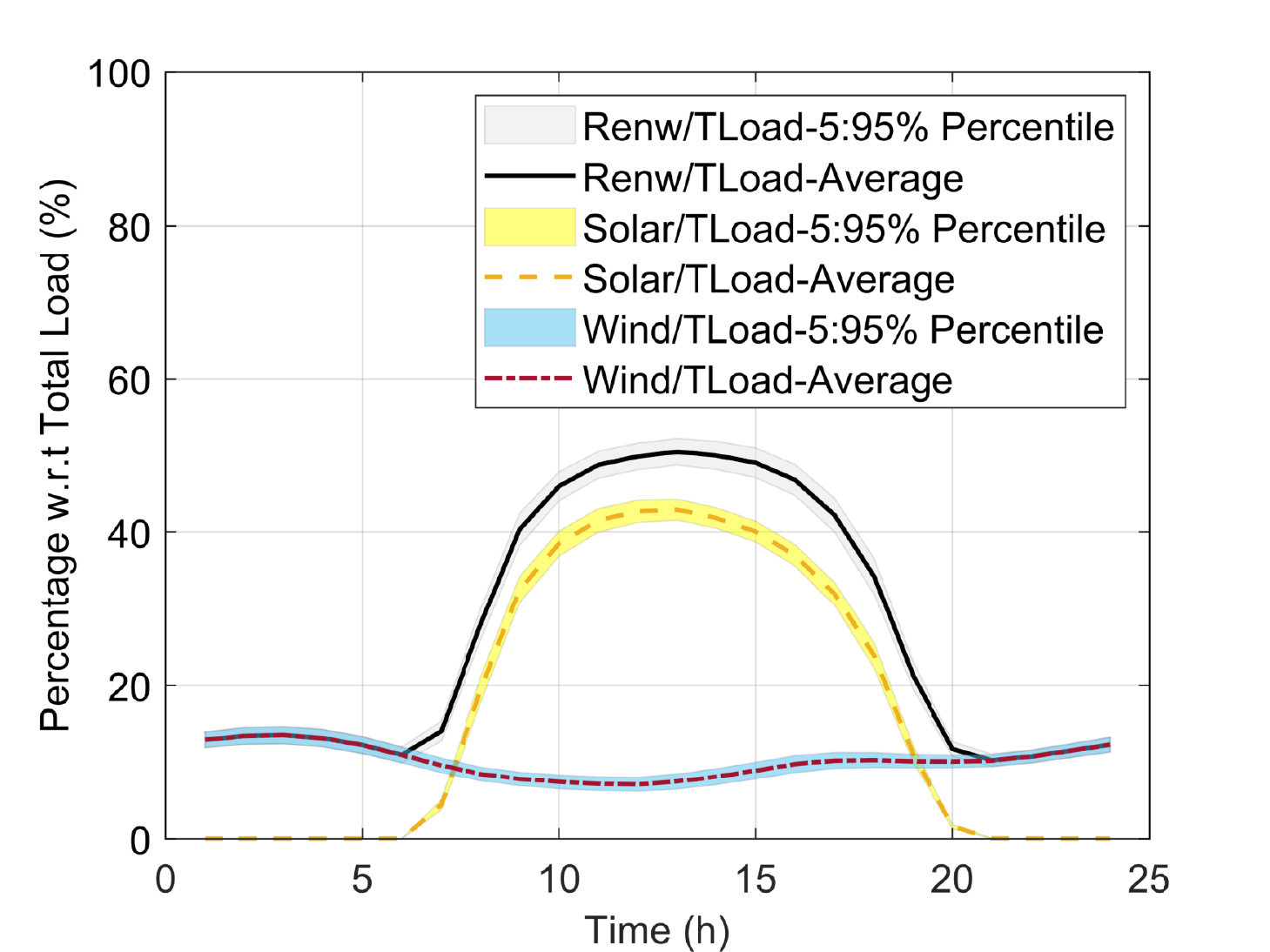} }

\caption{2020 renewable penetration with respect to total load in every U.S. RTO, 5\% to 95\% percentiles for: a) ERCOT, b) ISO-NE, c) MISO, d) NYISO, e) SPP, and f) CAISO. \textit{(PJM is also analyzed but excluded due to space limitation)}.} 
\label{fig:penetrations}  
\end{figure*}

The second analysis focuses on exploring which RTO had the highest penetration of RES, with respect to the total load, in the system during the 2020 COVID-19 outbreak. Fig. \ref{fig:penetrations} depicts the 5-95\% percentiles of RES penetration (both solar and wind) with respect to the total load of the specific RTO during the analyzed period. PJM is also analyzed but excluded in the figure due to space limitation and limited RES penetration. As seen in the figure, the RTO with the highest penetration of RES (mostly wind energy) is the SPP RTO. In this system, the RES penetration oscillates between 26\% and 47\% penetration. Other RTOs with high penetration of RES are CAISO, with a solar PV penetration of around 50\% and a wind energy penetration of around 18\%, and MISO with a wind energy penetration of around 19\%. 

\begin{figure}[t]
\centerline{\includegraphics[width=0.4\textwidth]{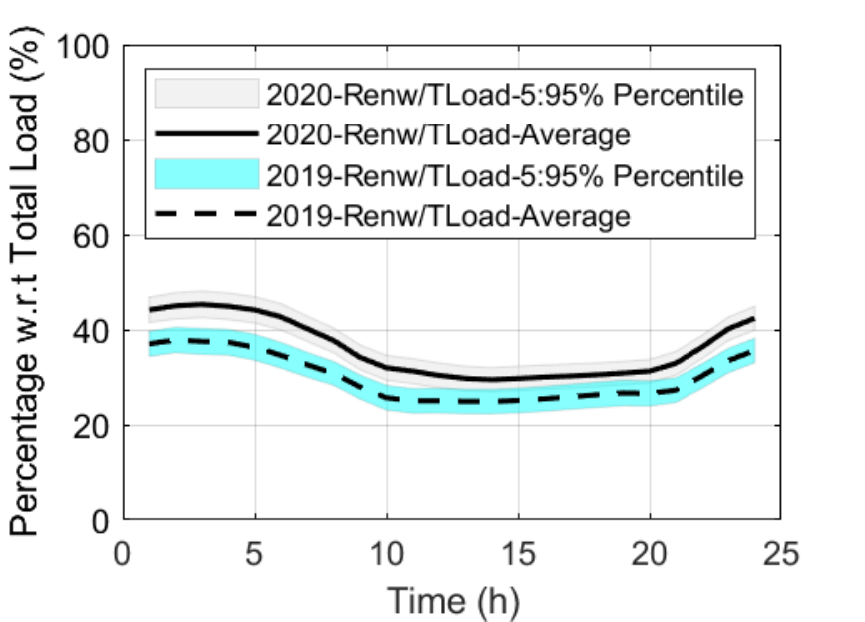}}
\caption{Renewable penetration with respect to (w.r.t) total load differences between 2019 and 2020 in SPP RTO.}
\label{fig:penetrations2019vs2020}
\vspace{-0.4 cm}
\end{figure}

From these analyses, we further analyze the differences in penetration between 2019 (normal year period) and 2020 (COVID-19 outbreak period) based on the load consumption reduction at various regions. The major differences in penetration of RES (between the two years) were found at the SPP RTO, where the wind penetration significantly increased, due to the load reduction, from approximately 37\% up to 45-47\%. Fig. \ref{fig:penetrations2019vs2020} shows the significant differences in penetration between the 2019 and the COVID-19 pandemic year (2020). As observed, the average pre-pandemic renewable penetration difference, in 24 hours, is approximately 8\%, making the system hit a maximum of 45-47\% of renewable wind energy penetration at some specific periods, thus making the system more vulnerable to possible LAAs with high impact in the system's frequency.

\vspace{-2mm}
\subsection{RTOs Frequency Requirements and Mitigation Procedures}

In the U.S., RTOs have very strict operational frequency thresholds for their systems. In this subsection, we examine the operational frequency thresholds for two different RTOs, ERCOT and NYISO, and for the nonprofit regulatory authority of the North American bulk power system, the North American Electric Reliability Corporation (NERC). Procedures taken by these operators to regulate overfrequency or underfrequency conditions are also discussed.

NERC requires systems to be operated within the 59.5 Hz (lower limit) and the 62.2 Hz (upper limit) frequency range \cite{NERC}. Protection relays must be put in place in order to protect the operational frequency of the system. These relays must trip whenever over-/underfrequency scenarios arise. The ERCOT RTO has similar frequency requirements for the activation of frequency protection mechanisms. The frequency thresholds for ERCOT are 59.3 Hz for underfrequency and 61.8 Hz for overfrequency \cite{guidessection}. Note that other protection mechanisms, such as load shedding, are also implemented to some extent by these operators according to complex rules designed to shed some percentage of the load in the system in order to abate frequency stability issues. NYISO is one of the RTOs that has stricter frequency requirements for both over and underfrequency control. The underfrequency threshold for NYISO is defined at 59.9 Hz while the overfrequency threshold is at 60.1 Hz. NYISO defines any event that causes the frequency to drop below or rise above the defined frequency thresholds as a `major disturbance'  \cite{NYISO}. Table \ref{tab:frequencythresholds} shows the specific threshold values (lower and upper-frequency limits) for ERCOT, NYISO, and NERC.

\begin{table}[t]
\centering
\caption{Operational frequency thresholds for ERCOT, NYISO, and NERC.}
\label{tab:frequencythresholds}
\begin{tabular}{||c|c|c|c||}
\hline\hline
\textbf{Operational Limits} & \textbf{NERC} & \textbf{ERCOT}& \textbf{NYISO}\\ \hline
Overfrequency (Hz)& 62.20 &61.80& 60.10 \\ \hline
Underfrequency (Hz)& 59.50 & 59.30 &59.90\\ 
 \hline\hline
\end{tabular}
\vspace{-4mm}
\end{table}

Generally, each RTO has its own set of mitigation procedures that are put in place to address any frequency excursions (over or underfrequency). However, these mitigation procedures tend to follow similar methodologies. Both NYISO and ERCOT follow similar underfrequency load shedding (UFLS) procedures where load shedding is enforced at consecutive load percentages according to the specific frequency thresholds. For instance, NYISO performs consecutive 7\% load shedding mechanisms when the frequency drops below 59.5 Hz, 59.3 Hz, 59.1 Hz, and 58.9 Hz respectively. If the frequency is still declining, the operator is required to take necessary actions that would minimize service interruption and equipment damage \cite{NYISO}. In ERCOT's side, the UFLS mechanism begins to be enforced when the frequency drops below 59.3 Hz by tripping 5\% of the total system load. If the frequency continues to decline, an additional 10\% load is shed at the 58.9 Hz threshold and a final additional 10\% of the load is shed at the 58.5 Hz limit \cite{ERCOT}. For overfrequency scenarios, most RTOs also follow similar procedures designed to be compliant with the NERC Balancing Authority Area Control Error (ACE) Limit (BAAL) standard. Sustained high frequencies are indications of major load-generation imbalances and if prolonged they can be considered as `major emergencies'. The procedure followed by most RTOs to wane overfrequency disturbances is the following \cite{NYISO}: (1) request over generating suppliers to adjust their generation, (2) reduce dispatchable generation to minimum operating limits, (3) request internal generators to operate in `manual' mode and below minimum dispatchable levels, (4) schedule variable load or storage to alleviate the overfrequency excursion, (5) reduce or cancel all transactions that are contributing to the imbalance, and (6) if the overfrequency event persists, declare a `major emergency' and de-commit generators until the imbalance is mitigated.

\section{Theoretical Analysis}
\label{sec:method}
In this section, we analyze the power grid dynamics under LAAs. 
In particular, our aim is to (1) determine the locations of the least-effort LAA within the power grid, (2) determine the amount of load to be manipulated under a LAA to cause unsafe frequency excursions, and (3) determine how actions (1) and (2) are affected by the low loading and reduced inertia conditions caused by COVID-19 lockdowns.  

In order to characterize these quantities, the system operator must determine how the system’s response  changes due  to the decrements in the power system inertia. While this can be predicted by conducting extensive simulations, such an analysis would be computationally expensive, since the operator must perform simulations under several inertia conditions and several combinations of nodes that could be subject to attack. In order to overcome this issue, our key idea is to compute the sensitivity of the system's eigensolutions (eigenvalues/vectors) with respect to the inertia of the system \cite{LakshIoT2021}, which in turn can be used to predict the change in the system's response due to RES penetration and the magnitude of LAAs that leads to unsafe events. We present the details of our analysis below.   

\vspace{-1mm}
\subsection{Power Grid Model Under Load-Altering Attacks}
We consider a generic power grid model consisting of $N$ buses connected by $M$ transmission lines. 
The set of buses are divided into generator and load buses, which we denote by $\mathcal{N}_G$ and $\mathcal{N}_L$, respectively. Let 
$N_G$ and $N_L$ denote the number of generator buses and  load buses, respectively. The power grid dynamics under LAAs can be modeled by a set of differential equations, given by \cite{AminiLAA2018}:
\begin{align}
& \begin{bmatrix} 
\Id & {\bf 0} & {\bf 0} \\
{\bf 0} & -\Mm & {\bf 0} \\
{\bf 0} & {\bf 0} & {\bf 0}
\end{bmatrix}
\begin{bmatrix} 
\dot{\deltav} \\
\dot{\omegav} \\
\dot{\thetav}
\end{bmatrix} = \begin{bmatrix} 
{\bf 0} \\
{\bf 0} \\
\pv^{LS} + \epsilonv^L
\end{bmatrix} + \nonumber \\
& \begin{bmatrix} 
{\bf 0} & \Id & {\bf 0} \\
\Km^I + \Bm^{GG} & \Km^P + \Dm^G  & \Bm^{GL} \\
\Bm^{LG} & {\bf O} & \Bm^{LL}
\end{bmatrix}
\begin{bmatrix} 
{\deltav} \\
{\omegav} \\
{\thetav} 
\end{bmatrix}, \label{eqn:dyn_mtx}
\end{align}
where $\deltav, \omegav \in \RR^{N_G}$ are the phase angle and rotor frequency deviations of the generator buses. The matrices $\Mm, \Dm^G,  \Km^I, \Km^P \in  \RR^{N_G \times N_G}$ are diagonal matrices whose diagonal entries are the generator inertia, damping, proportional, and integral coefficients, respectively. Matrices $\Bm^{GG} \in \RR^{N_G \times N_G}, \Bm^{LL} \in \RR^{N_L \times N_L}, \Bm^{GL} \in \RR^{N_G \times N_L}$ are sub-matrices of the admittance matrix, derived as
$\Bm_{bus} = \begin{bmatrix} \Bm^{GG}  & \Bm^{GL} \\ 
\Bm^{LG} & \Bm^{LL}
\end{bmatrix}.$ The vectors
$\pv^{LS}, \epsilonv^L \in \RR^{N_L}$ model the system load. Specifically, we assume that the total system load consists of two components, i.e., $\pv^L = \pv^{LS} + \pv^{LV},$ where $\pv^{LS}$ is the secure part of the system load (i.e., that includes non-smart and/or protected loads) and $\pv^{LV}$ is the vulnerable portion of the load. Under LAAs, the net load of the system is given by:
\begin{align}
   \pv^L =  \pv^{LS} + \epsilonv^L,
\end{align}
where $\epsilonv^L \in \RR^L$ is the LAA component, and $\epsilon_i \leq P^{LV}_i , i = 1,\dots,N_L$, where $\epsilon_i$ and $P^{LV}_i$ are the $i^{\text{th}}$ elements of the corresponding vector quantities. Next, we present theoretical results to analyze LAAs under COVID-19 low-inertia conditions.

\subsection{Analysis of LAAs Under COVID-19 Low-Inertia Conditions}
First, we characterize the power grid dynamics under LAAs, i.e., the solution to Eq. \eqref{eqn:dyn_mtx}.
Reference \cite{LakshIoT2021} derived closed-form expressions for the system response to LAAs using the theory of second-order dynamical systems. Assuming zero initial conditions, the response of the system due to a LAA vector $\epsilonv^L$ can be expressed as a function of the eigensolutions as: 
\begin{align}
{\bf \mathfrak z} (t)  = \sum^{N_L}_{i = 1} \epsilon^L_{i}  \fv_i(t), \label{eqn:response_load}
\end{align}
where 
\begin{align}
\fv_i(t) = \sum^{2N_G}_{j = 1} \LB {\frac{e^{\lambda_jt}-1}{\lambda_j}} \RB k_{ji} \zv_j. \label{eqn:perunit}
\end{align}
Note that ${\bf \mathfrak z} (t)$ and $\fv_i(t)$ are $2N_G-$dimensional vectors where the first $N_G$ elements represent the fluctuations of the generator phase angles $\deltav(t) \in \RR^{N_G}$ and the next $N_G$ elements represent the fluctuations of the generator phase angles $\omegav(t) \in \RR^{N_G}.$
In Eqs. \eqref{eqn:response_load} and \eqref{eqn:perunit},  $\lambda_j, \yv_j$, and $\zv_j$ are the eigenvalues and left and right eigenvectors of the system of differential equations described in Eq. \eqref{eqn:dyn_mtx}, which can be computed as a solution to the following equations: 
\begin{align}
\lambda_j \mathcal{A} \zv_j + \mathcal{B} \zv_j & = \b0,
\forall j=1,\cdots, 2{N}_G, \nonumber \\
\lambda_j \yv_j^\top \mathcal{A} + \yv_j^\top \mathcal{B} & = \b0, \forall j=1,\cdots, 2{N}_G. \label{eqn:eigenvals}
\end{align}
In Eq. \eqref{eqn:eigenvals}, the matrices  $\mathcal{A}$ and $\mathcal{B}$ are given by:  
\begin{align}
 \mathcal{A} &=
\begin{bmatrix}
 \Cm &  \Mm \\
 \Mm &  {\bf O}
\end{bmatrix},
\mathcal{B}  =
\begin{bmatrix}
 \Gm & {\bf O} \\
{\bf O} &  -  \Mm
\end{bmatrix}, \label{eqn:ABMtx}
\end{align}
\noindent where $\Gm = -(\Km^I + \Bm^{GG} - \Bm^{M} \Bm_{LG}),$
 $\Cm = -(\Km^P + \Dm^G),$ and $\Bm^M = \Bm^{GL} (\Bm^{LL})^{-1}.$ 
Furthermore, $\kv_j $ is a row vector given by $\kv_j  = \yv_j^{\top} \Bm^M \in \RR^{1 \times N_L}$ and $k_{ji}$ is the $i^{\text{th}}$ element of $\kv_j .$
Eqs. \eqref{eqn:response_load} and \eqref{eqn:perunit} can be interpreted as follows. First note that the response of the system ${\bf \mathfrak z} (t) $ is a linear function of the LAA vector $\epsilonv^L.$ In this context, $\fv_i(t)$ can be interpreted as the response of the system to a LAA of unit magnitude at the load bus $i \in \mathcal{N}_L.$ 

Let us consider that we are interested in the frequency fluctuation of the generator bus $n.$ Let $\omega^{max}_n$ denote the maximum frequency deviation (from the nominal grid frequency) beyond which safety mechanisms (such as load shedding etc.) are triggered. 
Then, from Eq. \eqref{eqn:response_load}, the minimum LAA at load bus $i$ that can cause an unsafe frequency excursion at the $n^\text{th}$ generator bus frequency can be computed as
\begin{align}
\epsilon^L_{{i},n} = \frac{\omega^{max}_n}{f_{i,n} (t^*_{i,n})}, i = 1,\dots,N_L, \label{eqn:least_load}
\end{align} 
where $t^*_{i,n} = \argmax_t f_{i,n}(t)$, where $f_{i,n}(t)$ is the $N_G+n^{\text{th}}$ component of the vector $\fv_i(t)$ (corresponding to the frequency component of the $n^\text{th}$ generator bus). Our objective is to investigate how the magnitude of the least-effort LAA  $\epsilon^L_{{i},n}$ changes due to COVID-19 low-load high-inertia conditions. 
To this end, we characterize the sensitivity of $\fv_{i}(t)$ with respect to the change in the inertia at generator node $g \in \mathcal{N}_G$. The result can be used to compute the change in the system's response due to LAAs under different levels of inertia, which can be subsequently used to determine the least-effort LAA that causes unsafe frequency excursions.

From Eq. \eqref{eqn:perunit}, note that $\fv_{i}(t)$ depends on the eigenvalues/vectors of the system.  
Thus, we must first characterize the sensitivity of the eigenvalues/vectors to changes in power system inertia in the following lemma.
\begin{lemma}
\label{prop:evsens}
 The sensitivity of the system's eigenvalue $\lambda_j$ with respect to the change of inertia at the generator node $g \in \mathcal{N_G},$ denoted by $\frac{\partial \lambda_j }{ \partial  M_g }$, can be computed analytically as
\begin{align}
\frac{\partial  \lambda_j}{\partial  M_{g}}   = -\lambda_j \yv^\top_j   \mathcal{I}_g  \zv_j,  j = 1,\dots,2N_G \label{eqn:sens_powergrid}
\end{align}
where \begin{align}
    \mathcal{I}_g = \begin{bmatrix}
{\bf O} & \lambda_j \Id_g \\
\lambda_j \Id_g  &  -\Id_g
\end{bmatrix},
\end{align}
and 
$\Id_{g} \in \RR^{N_G \times N_G}$ is a matrix whose $g^{\text{th}}$ diagonal entry is $1,$ and all other entries are zero. 

Furthermore, the sensitivity of left/right eigenvectors, denoted by $\frac{\partial \zv_j }{ \partial  M_g }$ and $\frac{\partial \yv_j }{ \partial  M_g }$ respectively, can be computed as
\begin{eqnarray}
\label{p4:eq2.3}
\frac{ \partial {\zv_j}} { \partial M_g} = \sum_{l=1}^{2{N_G}} a^{(\alpha)}_{jl} \zv_l 
\ \text{and} \
  \frac{ \partial {\yv_j}} { \partial M_g}  = \sum_{l=1}^{2{N_G}} b^{(\alpha)}_{jl} \yv_l,
\end{eqnarray}
where $a^{(\alpha)}_{jl}$ and $b^{(\alpha)}_{jl}$,  $\forall \,
l=1, \cdots ,2N_G$ are sets of complex constants defined as 
\begin{align*}
a^{(\alpha)}_{jl}  & =
- \yv_l^\top \mathcal{I}_g  \zv_j,
 l=1,\cdots,2{N}_G; l \neq j, \\
b^{(\alpha)}_{jl}  &=
- \yv_j^\top \mathcal{I}_g  \zv_l,
 l=1,\cdots,2{N}_G; l \neq j,
\end{align*}
and $\ a^{(\alpha)}_{jj} = b^{(\alpha)}_{jj} =
-  \lambda_j  \yv_j^\top \zv_j$.
\end{lemma}
The derivation of Lemma~\ref{prop:evsens} is presented in the Appendix. The sensitivity of $\fv_i(t)$ with respect to the change in the inertia of the system will, in turn, depend on the eigenvalue/vector sensitivities. The main result is presented in the following theorem.

\begin{theorem}
\label{lem:LAA}
The sensitivity of $\fv_i(t)$ with respect to the change of inertia at the generator node $g \in \mathcal{N_G},$ denoted by $\frac{\partial \fv_i(t) }{ \partial  M_g },$ can be computed analytically as
\begin{align}
\frac{\partial \fv_i(t) }{ \partial  M_g } & = \sum^{2N_G}_{j = 1} \left( \frac {\left(1 + \left[\lambda_j(t) -1\right] e^{\lambda_j(t)} \right)} {\lambda^2_j}
\derip{\lambda_j}
k_{ji} \right.   \nonumber
\\
&  \left. + { \LB \frac{e^{\lambda_j(t)}-1}{\lambda_j} \RB}
\left(\derip{\yv_j}\right)^T \bv^i_M  \right) \zv_j \nonumber \\ & + \sum^{2N_G}_{j = 1} \LB {\frac{e^{\lambda_jt}-1}{\lambda_j}} \RB k_{ji} \derip{\zv_j},
\end{align}
where $\bv^i_M$ is the $i^{th}$ column of the matrix $\Bm_M.$
\end{theorem}
The derivation of Theorem~\ref{lem:LAA} is presented in the Appendix. 

With slight abuse of notation, for a system with no renewable energy penetration, let us denote the value of $\fv_i(t)$ by $\fv^{0}_i(t).$ 
Assume that the inertia of the generator at node $g \in \mathcal{G}$ changes by $\Delta M_g.$
Then, using Lemma~\ref{lem:LAA}, $\fv_i(t)$ for the corresponding system can be approximated as
\begin{align}
\widehat{\fv}_i (t) =  \fv^{0}_i(t) + \sum_{g \in \mathcal{G}}\frac{\partial \fv_i(t) }{ \partial  M_g } \Delta M_g.
\end{align}
Let $\widehat{t}^*_{i,n} = \argmax_t \widehat{f}_{i,n}(t).$ Then, the least-effort LAA at load bus $i$ that causes an unsafe frequency deviation of the frequency of generator bus $n$ can be computed as
\begin{align}
\widehat{\epsilon}^L_{{i},n} = \frac{\omega^{max}_n}{ \widehat{f}_{i,n} (\widehat{t}^*_{i,n})}, i = 1,\dots,N_L, \label{eqn:least_load_1}
\end{align} 
The location of the least effort LAA is then given by $i^* = \arg \min_{i \in \mathcal{N}_L} \widehat{\epsilon}^L_{{i},n}.$

In the following section, we examine the validity of the sensitivity-based approach in predicting the magnitude and location of the least-effort LAA using simulations.

\subsection{LAA Formulation considering System Reserves}

{The LAA formulation is modified to consider system information related to the operating reserve margins available in the system.  In essence, the LAA is able to take advantage of the variability and uncertainty caused by renewable generation, by leveraging publicly available information related to operating reserve margins via the use of open-source intelligence (OSINT) techniques.}

{As explained in \cite{ela2011operating}, hourly scheduled generation needs to be complemented with operating reserves in order to ensure the balance of supply and demand in the minute-by-minute time scale. There are certain procedures that are set forth by different entities on how many operating reserves are required for the system. These procedures include information related to who can provide the reserves, when they need to be deployed, and how they are deployed. The reliability criteria to determine these procedures differ substantially from region to region, and many studies have shown that systems with high penetration of renewables need innovative methods and policies that can take into account increased variability and uncertainty \cite{ela2011operating}.The traditional scheduling and operating process of power systems are as follows:}

\begin{itemize}
    \item \textbf{Forward scheduling}: {schedules and unit commitment are obtained to meet the general load pattern of the day.}
    
    \item \textbf{Load following}: {within the day, the general trending load pattern is followed by performing economic dispatch and/or starting/stopping quick-start combustion turbines or hydro facilities.}
    
    \item \textbf{Regulation}: {the process of balancing fast second-to-second and minute-to-minute random variations in load or generation.}
\end{itemize}

\begin{figure}[h]
\centerline{\includegraphics[width=0.4\textwidth]{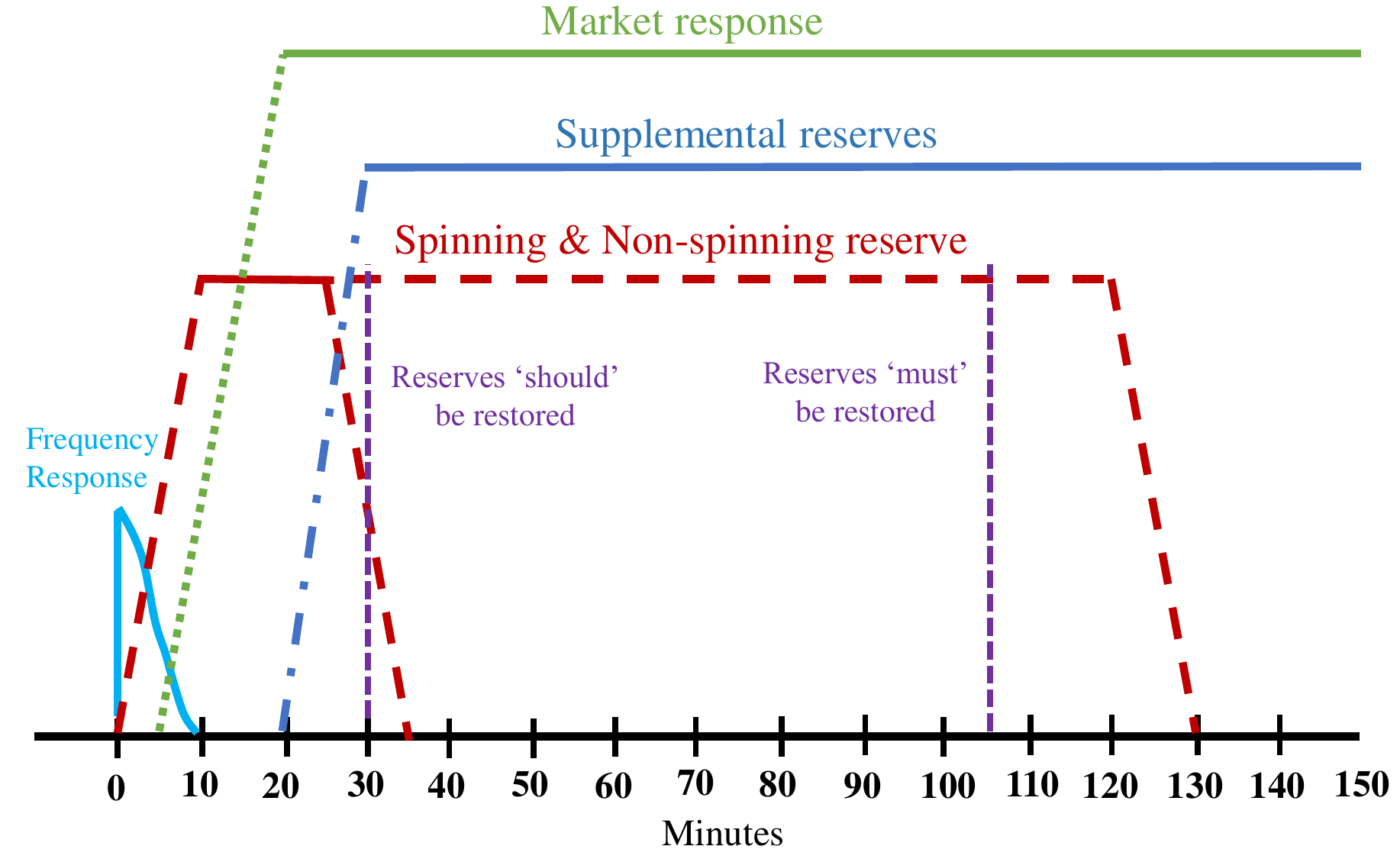}}
\caption{North American procedures for operating reserves.}
\label{fig:operatingreserves}
\end{figure}

{Fig. \ref{fig:operatingreserves} shows how operating reserves need to be used in a coordinated fashion to respond to emergency events according to North American procedures. These emergencies can be caused by natural events or cyber-related events. As seen in the figure, there are multiple responses, which vary by their length and response times, that are coordinated to address disturbances in the system. In most systems, the rules concerning the use of operating reserves are not fully defined, and may drastically vary from region to region. For instance, in the East Central Area Reliability Council (ECAR), the requirements for spinning and non-spinning reserve are both 3\% of daily peak load \cite{prada1999value}. Similarly, the Northwest Power Pool sets the reserve requirements to 3\% of load plus 3\% of generation, or to the magnitude of the single largest system component failure, whichever is larger \cite{power2016seventh}. In the mid-Atlantic region, the spinning reserve must be equal to the largest unit online. In Florida, spinning reserves must equal 25\% of the largest unit online. The Western Systems Coordinating Council requires operating reserves that are equal to 5\% of the load supplied by hydroelectric resources plus 7\% of the load supplied by traditional generation. These differences show that beyond NERC’s operating standard, there are no standardized adopted reserve requirements \cite{prada1999value, hummon2013fundamental}.}

\begin{figure}[h]
\centerline{\includegraphics[width=0.4\textwidth]{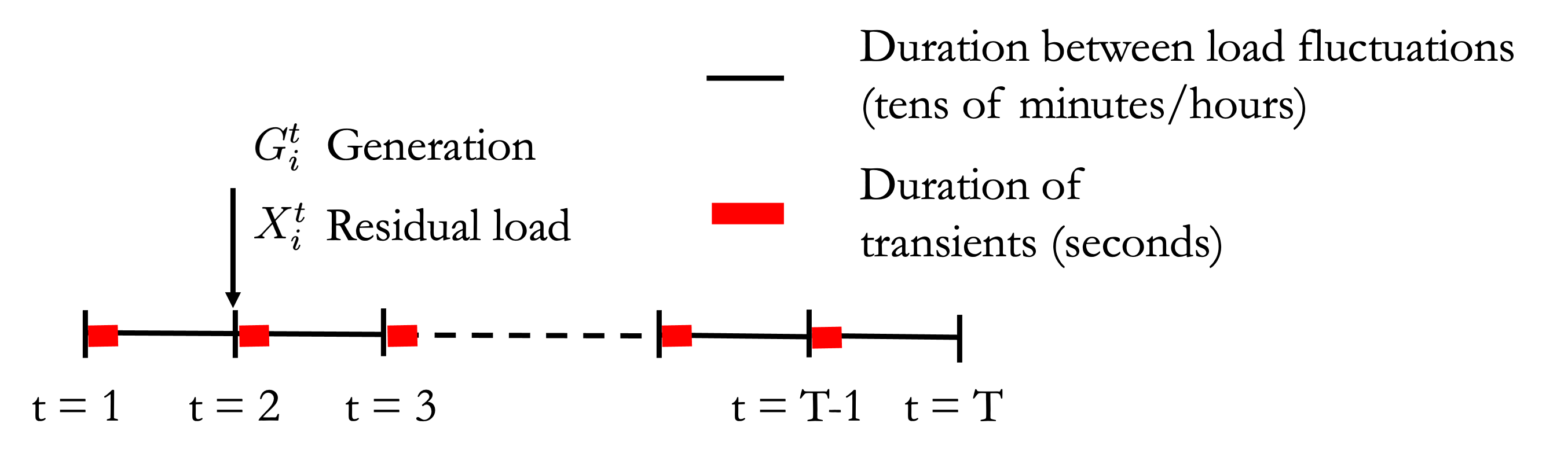}}
\caption{Timescale of load fluctuations and transients.}
\label{fig:Time}
\end{figure}

{Considering these differences in operating reserve margin requirements, the proposed LAA is modified to take into account reserve margins information that can be leveraged to perform a more targeted attack. With this knowledge (obtained using OSINT), the LAA can focus on compromising a large number of loads that would surpass the amount of operating reserves available in the system, and even other mechanisms, such as market response, would struggle to keep up.}

{The generator scheduling and operating reserves can be easily incorporated into the theoretical framework presented in Section~IV-B. Consider a slotted time setup indexed by $t = 1,2,\dots,T$ as shown in Fig.~\ref{fig:Time}. Herein, the time indices may, for instance, refer to different hours of the day. In comparison, the power grid transients only last for a few seconds as shown in the figure. 

Our objective is to analyze how the temporal fluctuations of the system load and generation affect the system's vulnerability to LAAs. To this end, we further introduce the following notations.
Let the residual load (i.e., the difference between the load and the renewable generation) at time $t$ be denoted by $X^t_i.$ We assume that the system operator obtains a forecast of the residual load, denoted by, $\widehat{X}^t_i.$ The operator schedules their conventional generation resources to balance the residual load, based on the forecasted value $\widehat{X}^t_i.$ At time $t,$ let $G^t_i$ denote the amount of generation scheduled. Furthermore, to account for contingencies, an operating reserve of  $R^t_i$ is maintained.

Note that if the residual load exceeds the scheduled generation plus the reserve margin (due to an inaccurate forecast), then the remaining load must be supplied by quick ramping generators. The mismatch can be modeled within the term $\pv^{LS}_i$ of \eqref{eqn:dyn_mtx} by setting $\pv^{LS}_i = \max(X^t_i - G^t_i - R^t_i, 0).$ Note that a non-zero value of $\pv^{LS}_i$ reduces the corresponding LAA $\epsilonv^L_i$ to cause unsafe frequency excursions. In Section \ref{sec:Sims}-C, we present simulation results to show how this mismatch affects the magnitude and time window of LAAs.

}

\section{Experimental Setup and Results}
\label{sec:Sims}
From the analyses conducted, it is clear that the COVID-19 pandemic has caused low demand conditions that, in turn, result in low-inertia grid conditions that could be exploited by attackers bent on causing harm to the electrical network through LAAs. In this section, theoretical and simulations results demonstrate the effects of possible LAAs in a power system with high penetration of RES and how the effort required (by the attacker) to trigger the emergency events changes due to such low-inertia conditions.

\begin{figure}[!t]
\centerline{\includegraphics[width=0.53\textwidth]{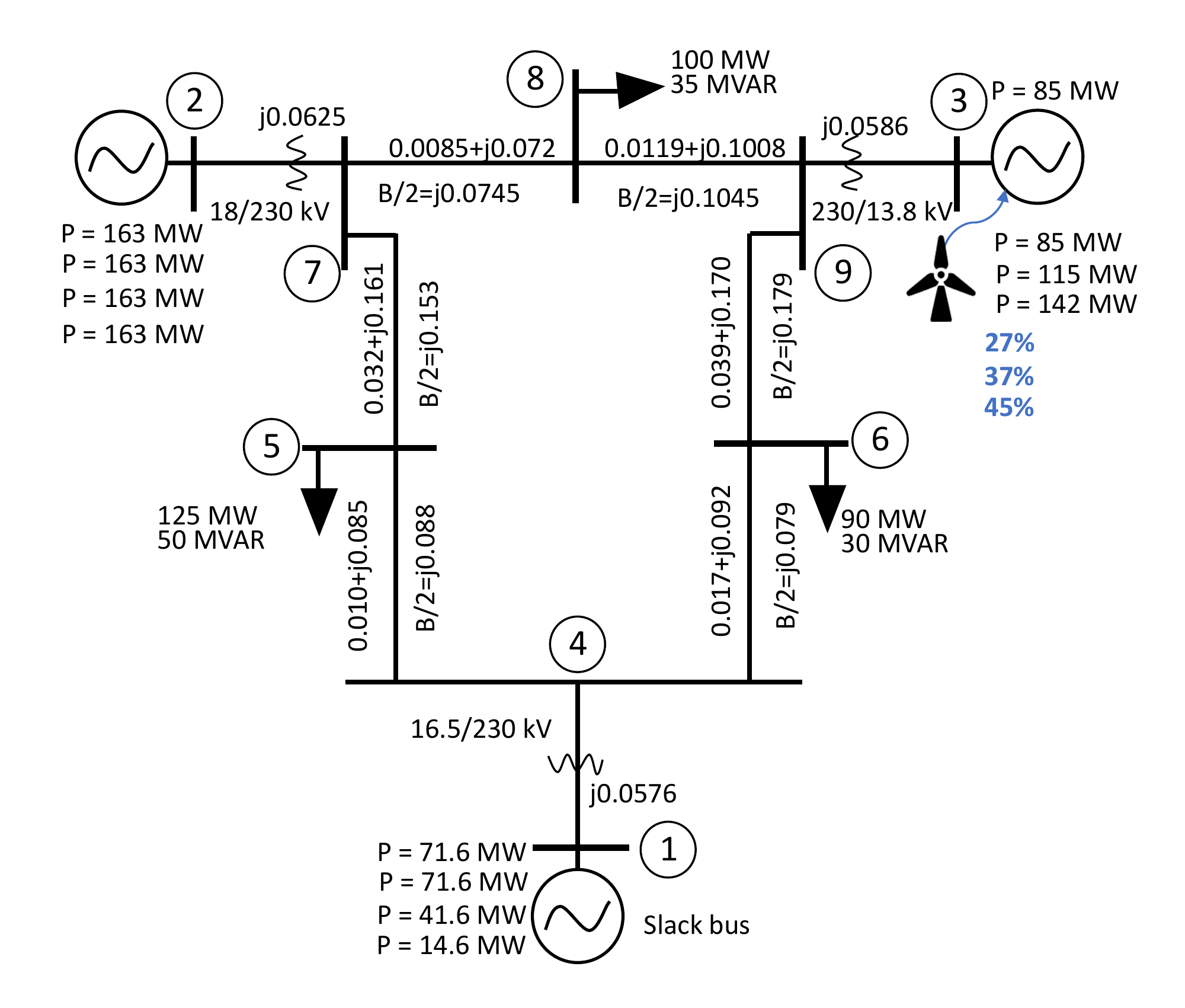}}
\vspace{-2mm}
\caption{WSCC 9-bus test system. The different active powers ($P$) shown represent the four different scenarios considered (i.e, $0 \%$, $27 \%$, $37 \%$, and $45 \%$ wind penetration).}
\vspace{-2mm}
\label{fig:wscc9bussystem}
\end{figure}

{In order to perform these analyses, the IEEE 118 bus test system and a modified version of the WSCC 9-bus system are used as test systems so that the impact of the proposed LAAs can be estimated. The IEEE 118 bus system is used as a motivating example that demonstrates the effects of high-impact LAAs. We consider the LAAs as propagating processes similarly to aggregated distribution system sections replaced by static or dynamic loads when simulating transmission system models \cite{kundur1994power, dvorkin2017iot, zografopoulos2021cyber}. In the WSCC 9-bus system modified system, generator \#3 is replaced with a wind generator composed of a wind turbine and a Doubly-Fed Induction Generator (DFIG). Various levels of wind penetration are considered. These are $0 \%$ (i.e., original system), $27 \%$, $37 \%$, and $45 \%$. These RES percentage penetrations mimic the RES penetrations observed in Section~II. Fig. \ref{fig:wscc9bussystem} shows the modified WSCC 9-bus system.}

\subsection{LAAs in IEEE 118 Bus System}

\begin{figure}[!t]
\centerline{\includegraphics[width=0.4\textwidth]{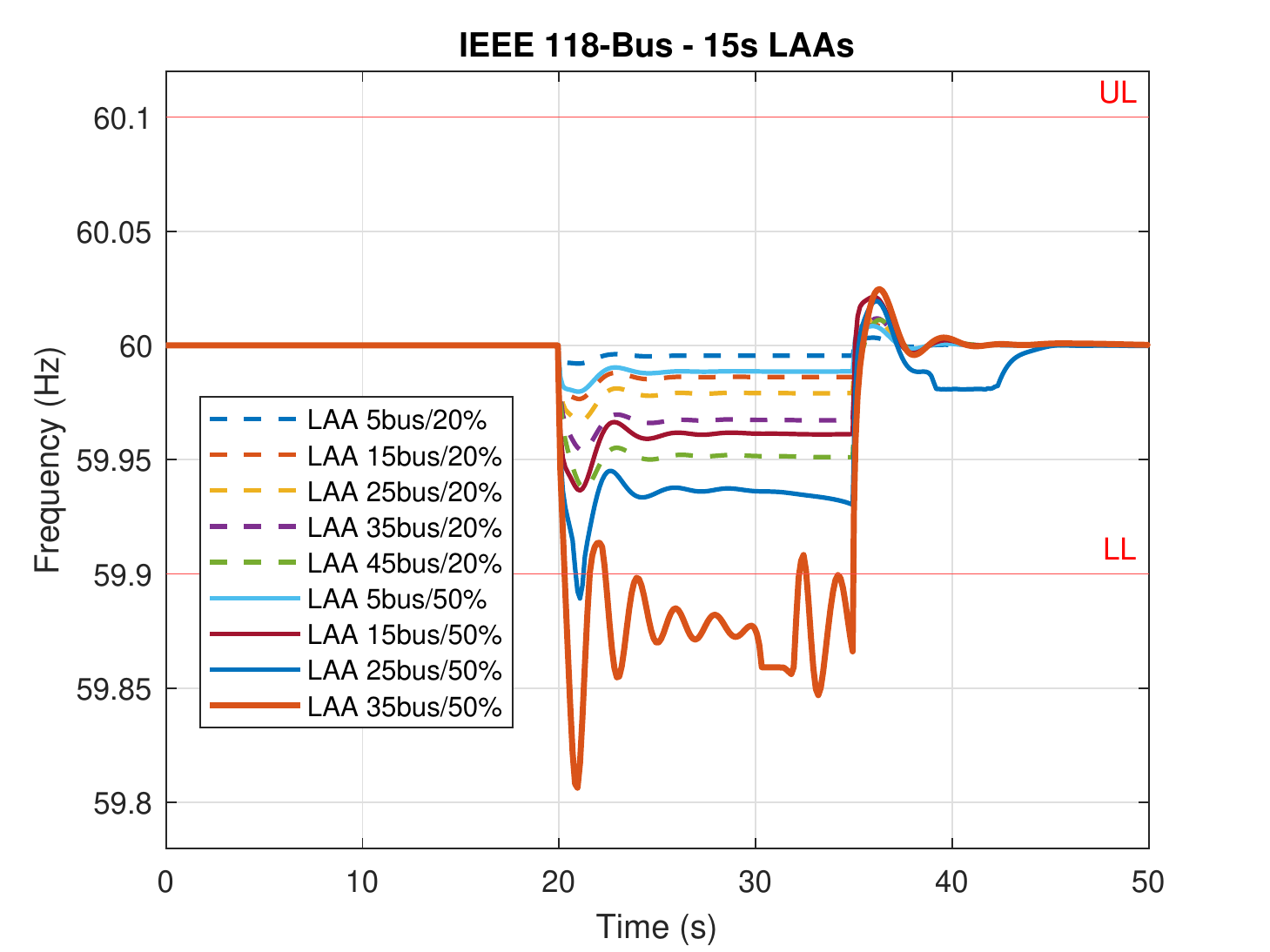}}
\caption{Frequency fluctuation caused by different LAAs targeted at multiple loads in the IEEE 118 bus system.}
\label{fig:118busfreq}
\end{figure}

{As a motivating example, we perform multiple LAAs on the original IEEE 118 bus test system with the objective of demonstrating the possible effects a LAA may cause in a large system. For these tests, LAAs are performed at two main compromised percentages: 20\% and 50\%, where the percentages indicate the percentage load increased at each compromised load bus for each case. Then, based on the specific compromised percentage, the number of loads compromised ranges from 5 loads up to 45 loads. Fig. \ref{fig:118busfreq} shows the frequency results obtained when multiple 15 seconds LAAs are deployed in the system. These frequency results are obtained by running multiple time-domain simulations of the IEEE 118 bus system using the Power System Analysis Toolbox (PSAT). As observed in the figure, severe LAAs, capable of increasing 50\% loading conditions of 25+ to 35+ buses (from a total of 91 load buses) in the system, have the capability of causing potential damage to the system by making it cross the lower frequency limit and causing instabilities. In turn, this state can be worse if the system in question has a high percentage of renewable penetration that can translate into low inertia conditions. For clarity purposes, a simpler test case, such as the WSCC-9 bus system, is used to explore the effects of LAAs in a dynamic system with low inertia conditions.}

\subsection{Results from Theoretical Analysis}

In this part, we present results from the theoretical analysis in Section \ref{sec:method}. 
The WSCC 9-bus system consists of three generator buses $\mathcal{G} = \{ 1,2,3 \}$ and six load buses $\mathcal{L} = \{ 4,5,6,7,8,9 \}.$  The power grid topological data is obtained from the MATPOWER simulator and the differential equations in Eq. \eqref{eqn:dyn_mtx} are simulated in MATLAB to obtain the generator dynamics under LAAs. 

\begin{figure}[!t]
\centerline{\includegraphics[width=0.45\textwidth]{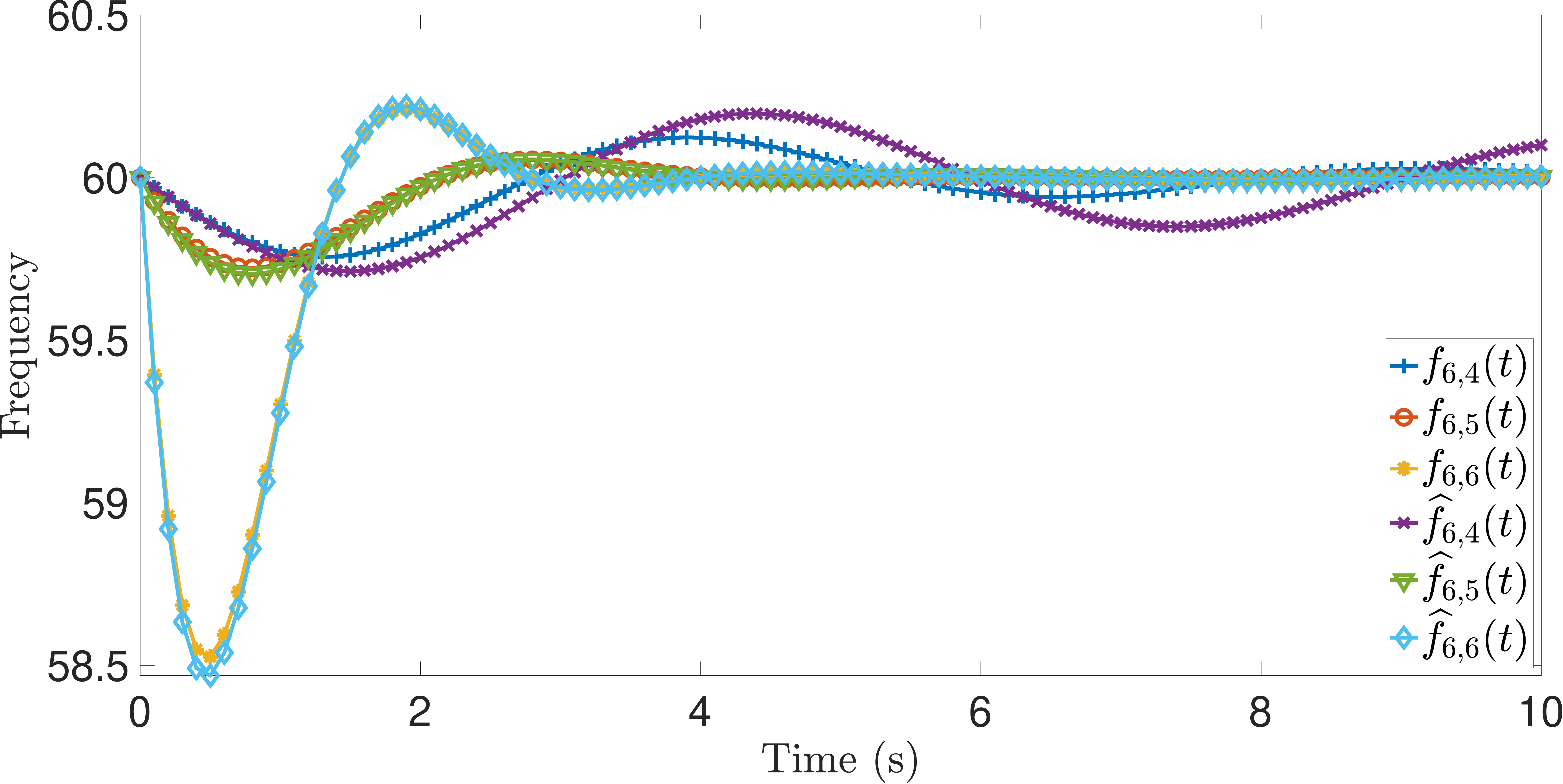}}
\caption{Frequency fluctuations $\{ f_{6,i}(t) \}^6_{i = 4}$ and their corresponding predictions based on the sensitivity approach $\{ \widehat{f}_{6,i}(t) \}^6_{i = 4}$ for 1~p.u. LAA injected at Bus~6. Simulations are performed using the WSCC~9 bus system with $45 \%$ wind penetration at generator Bus~3.}
\label{fig:perunit_fluc}
\end{figure}

First, we verify the validity of the sensitivity approach (presented in Lemma~\ref{lem:LAA}) in predicting the system dynamics under changing inertia conditions. 
As noted above, we consider $45 \%$ wind penetration at generator bus \#3 to mimic the power grid under COVID-19 low inertia conditions and inject a LAA at bus \#6. The fluctuations of $\{ {f}_{6,i}(t) \}^6_{i = 4}$ and $\{ \widehat{f}_{6,i}(t) \}^6_{i = 4}$ are plotted in Fig.~\ref{fig:perunit_fluc}. Note that these curves correspond to the frequency fluctuations at the generator buses and their predictions based on the sensitivity approach. The results show a close match between the curves, which confirms the validity of the theoretical analysis presented in Section~\ref{sec:method} to analyze LAAs under different levels of RES penetration levels.

We also compute the magnitude of the least-effort LAA that causes a frequency deviation of $0.1~$Hz considering different victim buses using the sensitivity-based approach. The values are enlisted in Table~\ref{tbl:LAA_Analytical}. 
We observe that victim bus \#6 is the location corresponding to the least-effort LAA. Furthermore, using the sensitivity approach, we can also predict the magnitude of the least-effort LAA that causes unsafe frequency fluctuations for different RES penetration levels in an efficient manner.

\begin{figure} \centering    
\subfigure[] { \label{fig:Load_Attack_Renew}     
\includegraphics[width=9cm]{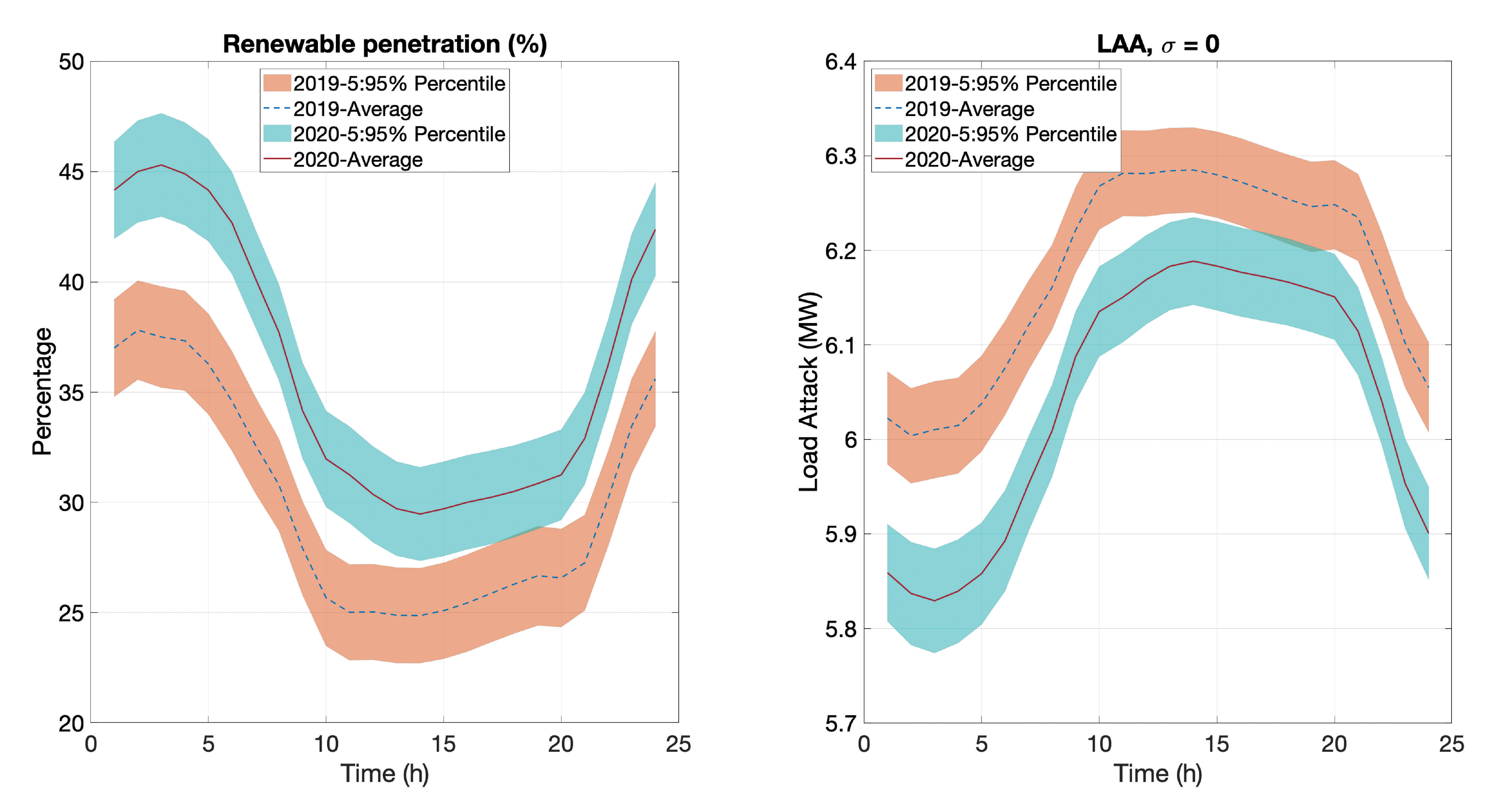}}
\subfigure[] { \label{fig:Load_Attack_Sch}     
\includegraphics[width=9cm]{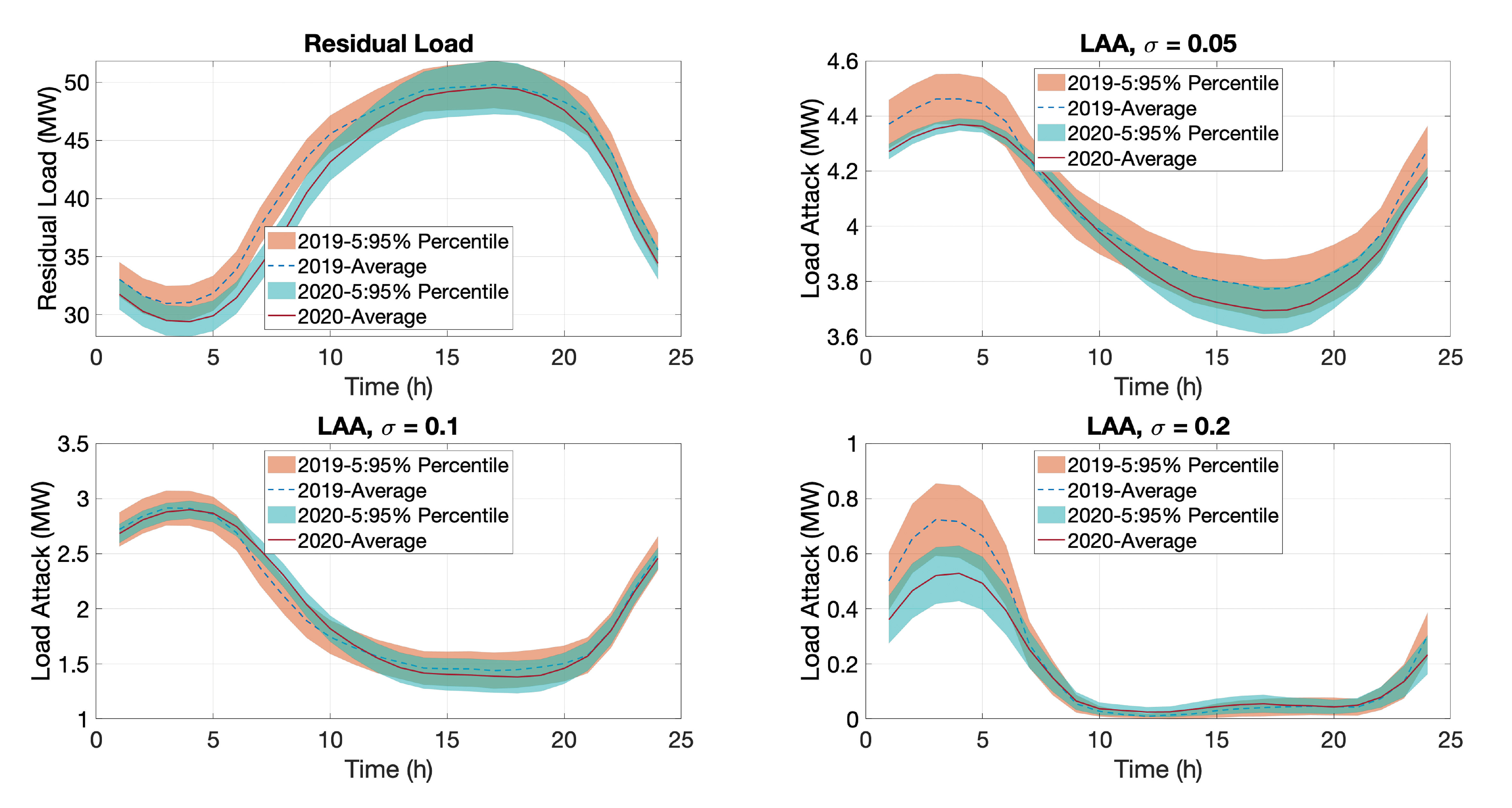}}

\caption{Magnitude of LAAs required to cause unsafe frequency excursions considering (a) temporal fluctuations of renewable energy (b) generator scheduling for SPP RTO.} 
\label{fig:Temporal}  
\end{figure}

\begin{table}[!t]
 \begin{center}
 \caption{The magnitude of the least-effort LAA in MWs that causes a frequency violation of $0.1~$ Hz for different renewable energy penetration scenarios. The simulations are performed using the WSCC-9 bus system.}
 \label{tbl:LAA_Analytical}
 \begin{tabular}{| c | c | c | c | c | c |} 
 \hline
 \hline
 Victim bus & $0 \%$ & $27 \%$ &  $37 \%$ &  $45 \%$ \\ [0.5ex] 
 \hline\hline
$4$ & $12.28$ & $12.28$ & $12.28$ &  $12.28$ \\ 
 \hline
$5$ & $14.15$ & $13.1$  &  $12.65$ &  $12.18$ \\
 \hline
{$6$} & {$6.77$} & {$6.26$}  & {$6.04$} &  {$5.81$}\\
 \hline
$7$ & $11.87$ & $10.99$  & $10.63$ &  $10.24$ \\
 \hline
$8$ & $10.2$ & $10.18$  & $10.17$ &  $10.17$ \\
\hline
$9$ & $16.58$ & $16.58$  & $16.58$ &  $16.58$ \\
\hline
\hline
\end{tabular}
\end{center}
\vspace{-3mm}
\end{table}  
{
\subsection{Impact of Renewable Energy Fluctuations and Generator Scheduling}
Next, we examine how the temporal fluctuations of renewable energy and the operator's generation scheduling impact the system's vulnerability to LAA. We use the IEEE-9 bus system to illustrate the results and inject LAAs at bus~6 (since it corresponds to the location of least-effort LAA). In order to mimic the real-world data, we consider the temporal fluctuation of renewable energy penetration from the SPP RTO as shown in the leftmost plot of Fig.~\ref{fig:Load_Attack_Sch}, and assume that an identical percentage of the load at bus~6 is served by renewable energy. We consider a generic residual load forecast model, given by  $\widehat{X}^t_i = X^t_i (1 + \sigma Y),$ where $Y \sim \mathcal{N} (0,1)$ is a Gaussian random number and $\sigma$ is the standard deviation of the forecast error as a fraction of the residual load.
 We simulate four scenarios by setting $\sigma$ to $0, 0.05, 0.1$ and $0.2$. We assume that generation is scheduled to match the forecast of the residual load, i.e., $G^t_i + R^t_i = \widehat{X}^t_i$ (note that the model can be trivially extended to the case when $G^t_i + R^t_i > \widehat{X}^t_i$).
 
The simulation results are presented in
in Figs.~\ref{fig:Load_Attack_Renew} and \ref{fig:Load_Attack_Sch}. The leftmost plot of Fig.~\ref{fig:Load_Attack_Sch} shows the temporal fluctuation of renewable energy. We observe that LAA required to cause unsafe frequency excursions is least when the penetration of renewable energy is high (hours $1-8$). This is expected since the inertia of the system is least at those times. Moreover, the LAA magnitude with $2020$ data is significantly less compared to the $2019$ data, due to higher renewable energy penetration in that year.

Next, we focus on the impact of generation scheduling in Fig.~\ref{fig:Load_Attack_Sch} by varying the standard deviation of the forecast error $\sigma.$ Recall that a higher value of $\sigma$ represents a higher mismatch between the scheduled generation and the residual load. Comparing Figs.~\ref{fig:Load_Attack_Renew} and \ref{fig:Load_Attack_Sch}, we clearly observe that scheduling mismatch makes the system more vulnerable to LAAs, as the mismatch must also be served by a fast ramping generator, similar to the LAA. Thus,
the attacker needs to inject a smaller amount of LAA to cause unsafe frequency excursions. Furthermore, in terms of the temporal variation, we observe that LAA magnitude decreases during the duration of the high residual load (hours $15-20$). This is because the forecast error is greater during these periods (recall that $\sigma$ is modeled as a percentage of the residual load). These results show that accurate forecast and accurate scheduling are essential to limit the time window of LAAs. 

}

\vspace{-2mm}
\subsection{Simulation Results Using PSAT}
The theoretical results presented in Section V-A are based on the second-order system model presented in Eq. \eqref{eqn:dyn_mtx}, which is based on some simplifications (such as the DC power flow model and a proportional–integral–derivative controller at the generators). In this subsection, we further validate the theoretical results using more realistic simulations based on PSAT with varying levels of wind energy penetration. As noted before, the wind generator is modeled as a wind turbine and a DFIG, while other generators are modeled as synchronous generators.

In order to analyze the effect of LAAs in the analyzed system (with different RES penetration), an 18 MW LAA that causes a 15-seconds load increase is deployed at different load buses in the system. Such an attack would correspond to manipulating approximately 36 commercial buildings with 50 AC units in each building (as presented in Section II), demonstrating the feasibility under a Botnet-type attack \cite{soltan2018blackiot}.

\begin{figure}[!t]
\centerline{\includegraphics[width=0.4\textwidth]{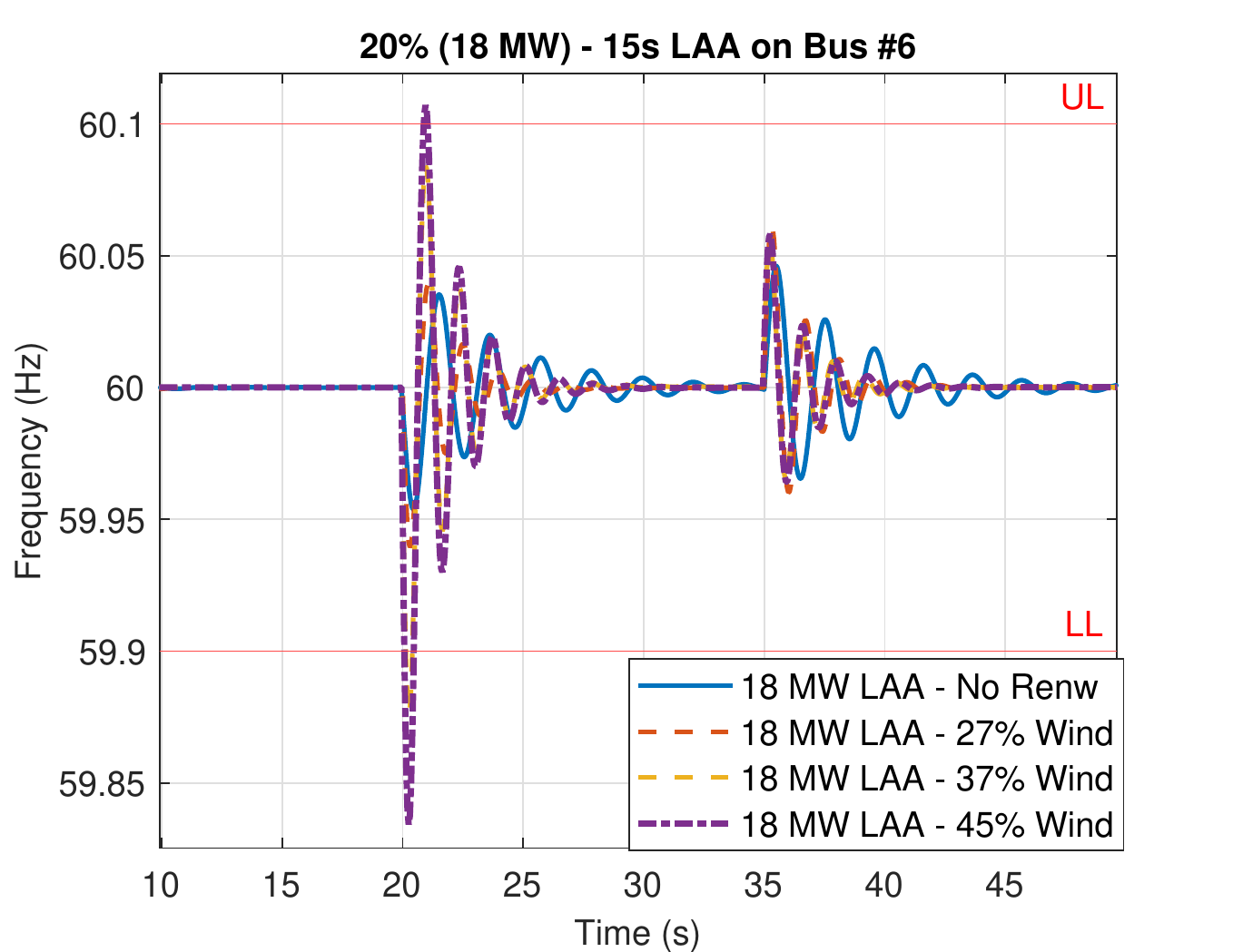}}
\caption{Frequency fluctuation caused by 18 MW LAA targeted at bus \#6. Different scenarios: a) No renewable (original system), b) 27\% wind penetration, c) 37\% wind penetration, and d) 45\% wind penetration.}
\label{fig:freq_plot_bus6}
\end{figure}

\begin{figure}[!t]
\centerline{\includegraphics[width=0.4\textwidth]{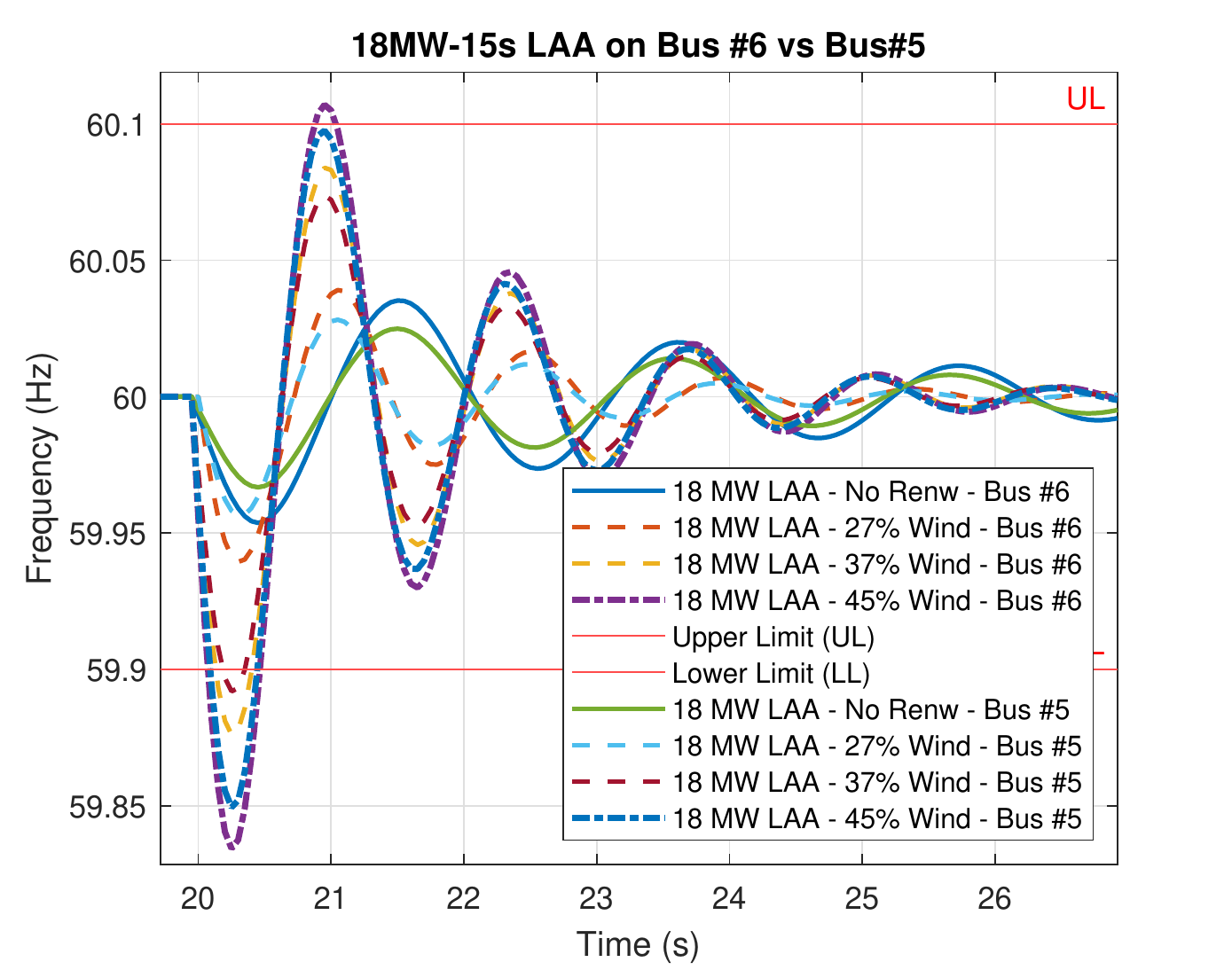}}
\caption{Frequency fluctuation caused by 18 MW LAA targeted at bus \#6 vs bus \#5. Different scenarios: a) No renewable (original system), b) 27\% wind penetration, c) 37\% wind penetration, and d) 45\% wind penetration.}
\label{fig:freq_plot_bus6vsbus5}
\end{figure}

Fig. \ref{fig:freq_plot_bus6} shows the frequency fluctuations in the different scenarios (i.e., $0 \%$, $27 \%$, $37 \%$, and $45 \%$) of the analyzed power system caused by a 15 seconds 18 MW LAA targeted at the most vulnerable bus in the system, i.e., bus \#6 (according to the previous analysis). As seen in this figure, the 18 MW LAA causes the 37\% and 45\% percent system scenarios to fluctuate out of the 0.1 Hz bounds set as the upper and lower frequency limits. Fig. \ref{fig:freq_plot_bus6vsbus5} shows a frequency graph that compares the frequency fluctuations caused by the 15 seconds -- 18 MW LAA targeted at bus \#5 and bus \#6 in the different analyzed scenarios. Here, it can be observed that the frequency fluctuations caused by the LAA targeted at bus \#6 are more prominent than the ones targeted at bus \#5. These results validate the results obtained by the theoretical analysis, where it was found that the most vulnerable bus in the system was bus \#6 since it requires the least-effort LAA.

The results of this section demonstrate that an identical attack can  lead to significantly larger frequency swings in low inertia grid conditions, which could cause unsafe frequency excursions, subsequently triggering generator trips and large-scale blackouts.

\vspace{-2mm}

\section{Conclusion \& Future Work}

In this article, we perform a comprehensive analysis of the impacts on electric power system operations caused by strict lockdown measures implemented worldwide during the COVID-19 pandemic. The implications of the COVID-19 mitigation measures, concerning power grid security, are analyzed using data compiled from seven different RTOs in the U.S., while the possibility and effects of a high-impact least-effort LAA are explored in low-inertia test system{s} that present high-penetration of renewable energy. Particularly, we present an extensive analysis related to the abnormal load reduction in different U.S. RTOs caused by COVID-19 lockdown measures. Based on this analysis, we formulate a theoretical formulation of a high-impact least-effort LAA targeted at high-wattage IoT-based devices. The formulated LAA is devised as a realistic cyber-attack capable of identifying the most vulnerable locations and amount of load needed to be compromised in order to cause unsafe frequency fluctuations in a low-inertia power system. Finally, we conduct theoretical and simulation-based experiments to demonstrate the effects of the formulated LAA in a low inertia system with a high penetration of RES, which are similar to the conditions that exist in one of the RTOs analyzed during the COVID-19 pandemic. {
Our results show that low-inertia conditions caused due to high penetration of renewable energy can make the grid vulnerable to LAAs. However, accurate forecast of renewable energy and accurate generation scheduling can significantly increase the effort required by the attacker to execute LAAs and limit the time window for such attacks. }

Future work will focus on exploring targeted LAAs in larger transmission and distribution systems with high-penetration of renewable energy, and 
{the exploration and development of defense strategies designed to enhance the grid's resilience in the face of COVID-19 type {of} events. Some of the defense strategies that will be explored are based on the implementation of security reinforcements, utilization of on-demand backup generators, security controls against misinformation campaigns, and the integration of optimally scheduled portable battery energy storage systems. Specifically, on the latter strategy, we plan to design a spatio-temporal decision model supporting load-leveling while addressing the congestion of potential LAAs. Another area that will be explored in future work is an analysis under a generalized non-linear power grid model encompassing multiple control areas. Finally, the development and deployment of a realistic software-based LAA that can be used to explore its cyber and physical effects in a system modeled in a hardware-in-the-loop (HIL) co-simulation platform will be important.}

\IEEEpeerreviewmaketitle

\bibliographystyle{IEEEtran} 
\bibliography{biblio}

\section*{Appendix}
In this Appendix, we present a sketch of the derivation of Lemma~\ref{prop:evsens} and Theorem~\ref{lem:LAA} respectively. First, we derive the parametric sensitivity of the eigenvalues with respect to the system inertia. The sensitivity of the eigenvalue of a second-order system with respect to the parameter $\alpha_m$ is given by \cite{jp14}:
\begin{equation}
\frac{ \partial{\lambda_j}}{ \partial \alpha_m} 
 = -
\yv_j^T\left [ \lambda_j  \frac{ \partial \mathcal{A}}{ \partial \alpha_m}  + \frac{ \partial\mathcal{B}}{ \partial \alpha_m}  \right] \zv_j. \label{eqn:sens_general}
\end{equation}
For matrices $\mathcal{A}$ and $\mathcal{B}$ defined in Eq. \eqref{eqn:ABMtx}, we have:
\begin{align}
    \frac{ \partial \mathcal{A}}{ \partial M_g} =  \begin{bmatrix}
{\bf O} &  \Id_g \\
 \Id_g  &  {\bf O}
\end{bmatrix},  \frac{ \partial \mathcal{B}}{ \partial M_g} = \begin{bmatrix}
{\bf O} &  {\bf O} \\
 {\bf O}  &  -\Id_g
\end{bmatrix}. \label{eqn:der_mtx}
\end{align}
Substituting Eq. \eqref{eqn:der_mtx} in Eq. \eqref{eqn:sens_general}, we can obtain the expression corresponding to the eigenvalue sensitivity.
The sensitivity of the eigenvectors can also be derived similarly and omitted due to the lack of space. 

Finally, we derive the sensitivity of $\fv_i(t)$ with respect to system inertia. To this end, we differentiate Eq. \eqref{eqn:perunit} with respect to $M_g, g \in \mathcal{G}$. First, using chain rule for derivatives, the derivative of $\fv_i(t)$ with respect to $M_g$ can be expressed as:
\begin{equation} 
\derip{ \fv_i (t)} = 
\sum_{j=1}^{2N_G}  \LB \derip{a_j(t)}\zv_j +  a_j(t) \derip{\zv_j} \RB, \label{eqn:der1}
\end{equation}
where $a_{ji}(t) = \LB {\frac{e^{\lambda_jt}-1}{\lambda_j}} \RB k_{ji}$ is a short-hand notation. Furthermore, using chain rule again, we can express $\derip{a_{ji}(t)}$ as:
\begin{align}
\derip{a_{ji}(t)}  =
\frac {\left(1 + \left[\lambda_j(t) -1\right] e^{\lambda_j(t)} \right)} {\lambda^2_j}
\derip{\lambda_j}
k_{ji}  \nonumber
\\
+{ \LB \frac{e^{\lambda_j(t)}-1}{\lambda_j} \RB}
\left(\derip{\yv_j}\right)^T \bv^i_M, \label{eqn:der2}
\end{align}
where $\bv^i_M$ is the $i^{th}$ column of the matrix $\Bm_M.$

Note that in Eqs. \eqref{eqn:der1} and \eqref{eqn:der2}, $\derip{\lambda_j}$ and $\derip{\zv_j}$ are the derivatives of eigenvalue and eigenvector  respectively with the system inertia. 

Combining Eqs. \eqref{eqn:der1}, \eqref{eqn:der2},  \eqref{eqn:sens_powergrid} and \eqref{p4:eq2.3}, we obtain the final result of Theorem~\ref{lem:LAA}.


\end{document}